\documentclass[oribibl]{llncs}
\usepackage{graphicx}
\usepackage{amsmath}
\usepackage{amssymb}
\usepackage{mathtools}
\usepackage{subfigure}
\usepackage{wrapfig}
\usepackage[T1]{fontenc}
\usepackage{lmodern}
\usepackage{color}

\usepackage{fancyhdr}

\begin{document}
\title{An extensible formal semantics for UML activity diagrams\thanks{Research supported by NSF Grant CCF-0926194. This information is subject to US-Export Controlled - ECCN:EAR-99} }
\author{Zamira Daw \and Rance Cleaveland}
\institute{Department of Computer Science, University of Maryland}

\maketitle

\begin{abstract}
This paper presents an operational semantics for UML activity diagrams. The purpose of this semantics is three-fold: to give a robust basis for verifying model correctness; to help validate model transformations; and to provide a well-formed basis for assessing whether a proposed extension/interpretation of the modeling language is consistent with the standard. The challenges of a general formal framework for UML models include the semi-formality of the semantics specification, the extensibility of the language, and (sometimes deliberate, sometimes accidental) under-specification of model behavior in the standard. Our approach is based on structural operational semantics, which can be extended according to domain-specific needs. The presented semantics has been implemented and tested. \end{abstract}

\section{Introduction}
Model-driven development (MDD) emphasizes the use of models and model transformations through the system development process. Automatic model transformations may be used to generate artifacts for implementation (e.g. code) or for analysis purposes (e.g. intermediate graph representations). This has increased the importance of verification methods for models and model transformations in order to ensure a correct development process. The Unified Modeling Language (UML) \cite{UML241} has attracted substantial attention as a language for MDD. UML is a non-proprietary, independently maintained standard that includes several graphical sublanguages and a precise abstract syntax given via a metamodel.  The semantics is given informally in a natural language, although for some UML sublanguages reference is made to more mathematical models such as state machines and Petri nets. UML also provides an extension mechanism that allows its adaptation to specific application domains.

This work focuses on \emph{UML activity diagrams}, which are generally used to specify the workflow of a system.  Activity diagrams can be seen as so-called block diagrams, with a system represented in terms of \emph{actions} (blocks) that compute outputs in terms of inputs, and edges and special-purpose nodes that together determine how data is routed from one action to another. The activity semantics is based on Petri nets semantics. Activity diagrams are among the most widely used behavioral diagrams in the UML standard, and find application in a variety of domains, including business-processes engineering~\cite{Cao06,Lam07,Eshuis06,Guelfi05}, embedded systems \cite{Grobelna10,Daw13}, and medical devices \cite{Daw2014}.

\begin{figure}[h]
\centering
 \includegraphics[width=0.6\textwidth]{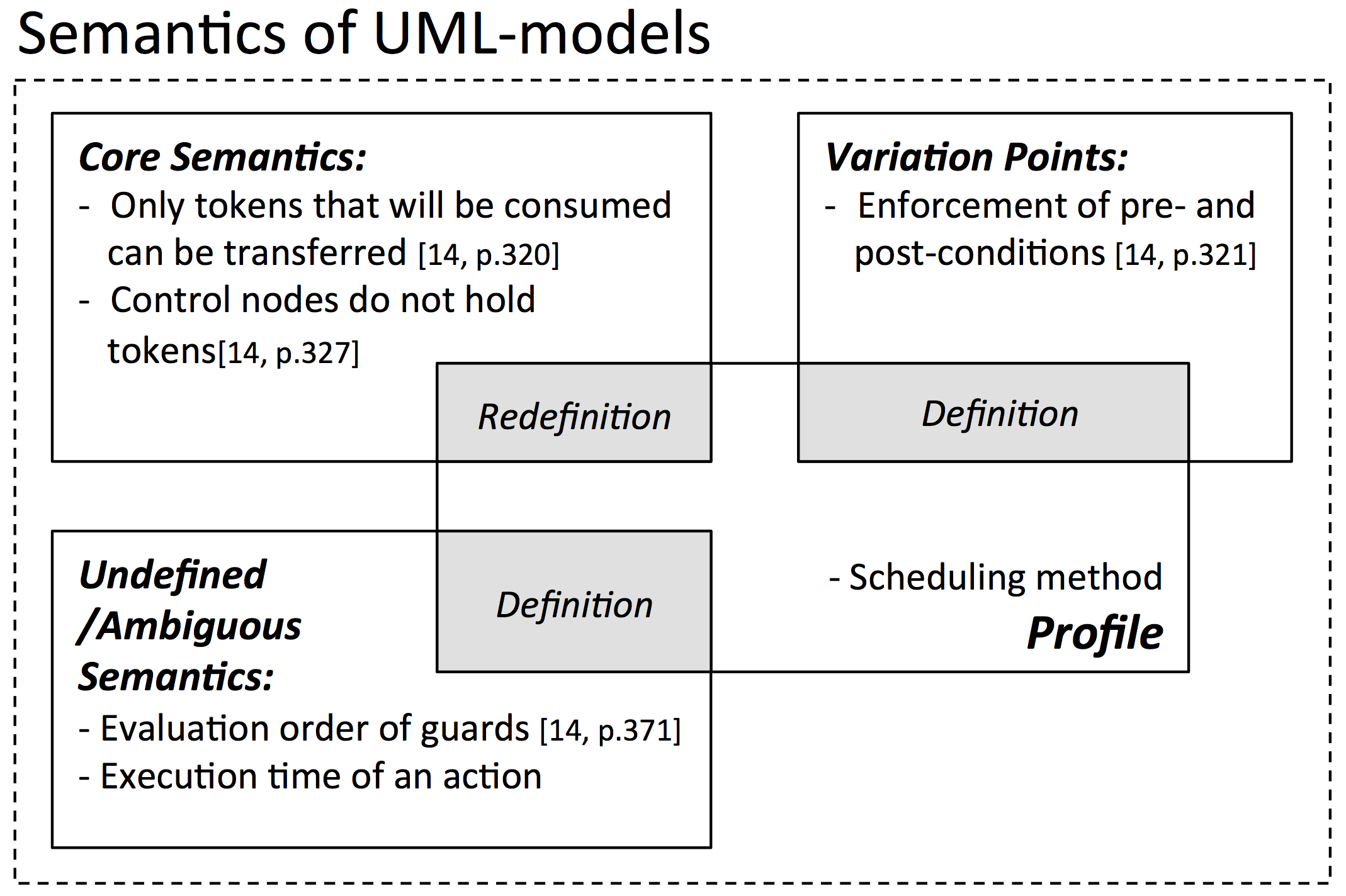} \label{fig:UMLsemantics}
 \caption{Semantic components of an interpretation of UML models.  Note that profile semantics may redefine the behavior of core concepts, make choices among semantic variation points, and resolve under-specification, as well as introduce new semantic concepts (e.g.\/ scheduling, priority).}
\end{figure}

The purpose of this paper\footnote{The work is part of a larger effort to formalize UML model semantics in a way that supports the extensibility and flexibility of the standard.} is to present a mathematically well-defined operational semantics for UML activity diagrams.  The research is motivated by several concerns.  On one hand, to realize the full benefit of MDD, engineers need mechanisms for checking the correctness of their models.  Formalizing these checks (e.g.\/ by using model checking) requires a mathematical account of the behavior of diagrams.  Reasoning about the correctness of model transformations (e.g.\/ code generation) also requires a precise account of model behavior in order to determine if the transformation correctly preserves model semantics.  At the same time, by design, UML may be interpreted flexibly~\cite{Broy11,Gulan13} and may be extended via profiles.  Determining when an interpretation/extension of UML is semantically consistent with the standard is challenging in the absence of a reference semantics.  The challenges are primarily attributable to the following characteristics of the standard:  ambiguities, under-specifications, semantic variation points, and semantic extensions using profiling. Figure~\ref{fig:UMLsemantics} categorizes the characteristics and gives examples related to activity diagrams.  It should be noted that some of the resulting semantic choices are explicitly identified in the standard; others are due to incomplete and sometimes contradictory exposition.

\vspace*{-0.4cm}
\vspace*{-0.4cm}

Our proposed reference semantics is a structural operational semantics (SOS) (Section 4).  SOS is chosen due to its implicit definition translation from models into Kripke structures, which are used in model-verification tools such as model checkers. The proposed semantics uses non-determinism to capture all possible behaviors in case of under-specification or ambiguity in the standard.  For example, the standard does not define the duration of the execution of an action as is shown in Figure \ref{fig:UMLsemantics}. In our semantics, the termination of an action is defined by an inference rule that is applicable after the invocation of the action but is non-deterministically executed with respect to the other applicable rules. Allowing all possible behaviors in this fashion ensures that the proposed semantics covers interpretations consistent with the standard, and that all possible behaviors can be taking into account during the verification. The semantics has been implemented using java language and tested using common activity model structures. Alternative domain-specific semantics can be defined by using the proposed rules, either explicitly or in a customized way, or by adding additional rules (Section 5).


\vspace{-0.2cm}

\section{Related Work}

\vspace{-0.2cm}
The work in this paper was initially inspired by research conducted by the authors and another collaborator on analyzing the performance of different model checkers on activity diagrams for medical-device systems~\cite{Daw13,Daw2014}.  That work required the implementation of translators from diagrams into the (formally precise) input notations of these tools.  Other researchers have also confronted this translation problem; examples include~\cite{Cao06,Grobelna10,Xu08,Lam07,Guelfi05,Eshuis06}.  Each of these pieces of work indirectly defines a formal semantics for activities via the translation strategies used.  An immediate question that emerges is this:  in what sense are these semantics correct, or consistent, with respect to the standard?  Providing a means for answering this question is a prime motivation for this paper.

For similar reasons, other researchers have also developed mathematical semantics for (fragments of) UML activity diagrams.  A formalization of the earlier  UML 1.x standard has been undertaken using Abstract State Machines (ASMs) as a basis in \cite{Borger00b}.  The more current 2.x standard has been given a semantics in terms of Petri Nets (PN) in~\cite{Storrle04,Staines08}.  Both approaches may be seen as translational in the sense that the accounts describe how activity diagrams may be interpreted as ASMs / PNs.  Another translational approach for UML 2.x is given in~\cite{Lano09}; that work interprets activity diagrams by giving a method for converting them into UML state machines.  It also gives another account of these diagrams in the action-semantics style of Mosses and Watt.  None of these approaches addresses the issue of extensibility nor discusses how to assess an extended semantics against a reference once.  They all omit certain aspects of activity-diagram behavior. 
Among these approaches, the PN-based semantics of St\"{o}rrle arguably has the broadest coverage of the standard by addressing data flow \cite{Storrle05b}, structured nodes \cite{Storrle04}, streaming  and exceptions \cite{Storrle05}. This semantics however does omit some aspects of the standard, such as: token-holding by all nodes, token-transfer limitations, and different types of invocations of activities, which are addressed in our semantics. St\"{o}rrle also points out some challenges of using a PN-based semantics, such as, the modeling of non-local behaviors (e.g. where tokens of an entire activity or a region have to be terminated, which is presented in final nodes or exceptions), which we believe are easier to address using SOS. In addition, model checkers based on Kripke structures have better capabilities than PN model checkers. It should be noted that one benefit of a PN semantics is the capability to model so-called \emph{true concurrency}, in which different model behaviors can happen simultaneously.  Such features can be captured in SOS also, although we do not do so in this paper for reasons of brevity; instead, our semantics reduces concurrency to interleaving.


Groenniger \cite{Groenniger10} and Knieke \cite{Knieke12} present a flexible semantics for a basic sub set of UML activities. Groenniger \cite{Groenniger10} proposes a denotational inner semantics with variation points. The inner semantics specifies the execution order of actions by using invariant definitions. The type of implementation of actions is defined using variation points (e.g. actions as methods).  Our approach allows customizing the implementation of  both actions and  execution order. Furthermore, the customization range of Groenniger's approach is limited to its target-domain, which is an object-oriented modeling language. Knieke \cite{Knieke12} presents a framework, which enables composition of operational semantics out of fundamental semantic constructs. These constructs define the state and the execution sequence of an activity diagram. The sequence is specified by a step algorithm, which is triggered by a global clock. This algorithm manages the activation of steps that consume or offer tokens. The steps can be customized according to the domain-specific needs. In contrast to our approach, Knieke uses a global clock and a step algorithm that synchronize execution nodes, which is not defined in the standard. Furthermore, the synchronization reduces the number of possible behaviors that can be specified, thereby limiting the set of target-domains. In addition, our approach enables the consistency verification of the extension with the UML standard.

\vspace{-0.2cm}
\section{Structure of UML activities}
\label{sec:Structure}
\vspace{-0.1cm}
The UML activity diagram is a graphical representation of control and data flow.  An activity contains nodes and edges. A node represents a function that takes a set of inputs and converts them into a set of outputs. An edge models the connection between two nodes. The execution of a node is determined by its connections and its constraints. 

\vspace{-0.5cm}
\begin{table} [!h] \caption{Definition of the structure of activity elements} \label{table:Structure}\begin{center}
\vspace{-0.5cm}
\begin{tabular}{l|l}
\textbf{Element type} & \textbf{Definition} \\
\hline 
Activity &$\langle  N, E, APN, PS \rangle$\\
Action & $\langle I, O,  m_{io} \rangle$  \\
CallBehaviorAction & $\langle I, O, behavior, synchronous  \rangle$ \\
Fork & $ \langle  i, O \rangle$\\
Edge & $\langle source, target, guard, weight \rangle$\\
Pin & $\langle direction, \delta, upperbound, upper, lower,  ordering \rangle$\\
ActivityParameterNode & $\langle direction, \delta, upperbound, upper, lower,  ordering, streaming, $\\
&$exception \rangle$
\end{tabular}
\end{center}
\end{table}

\vspace{-1cm}
The structure of activities is defined in the standard by a metamodel, which specifies type of elements,  type of connections between elements and elements' properties  that a diagram can have. Table \ref{table:Structure} summarizes the properties associated to activity elements that are used in this paper.

An \textbf{\textit{activity}} is composed by a set of nodes $N$, a set of edges $E$, and a set of activity parameter nodes $APN$, which can be grouped in parameter sets  $PS$ ($PS \subseteq 2^{APN}$). $APN$ represent the interface of an activity. $Action$, $CallBehaviorAction$ and $Fork$ are node types. An \textit{\textbf{action}} is the fundamental unit of executable functionality of an activity.  An \textit{action} represents a functionality that maps a set of inputs $I$ into a set of outputs $O$. This mapping is defined outside of the model and represented as $m_{io}:  n \rightarrow ((inpin(n)\rightarrow D^*) \rightarrow (outpin(n) \rightarrow D^*))$. A \textit{\textbf{CallBehaviorAction}} invokes a behavior defined by the activity $behavior$. Using this type of node, a system can be built using multiple hierarchical levels. The invocation can be $synchronous$ forcing the action to wait until the termination of the behavior's execution. Otherwise, the execution of the action terminates with the invocation of the behavior. A \textit{\textbf{Fork}} is a control node that splits a flow into multiple concurrent flows. This node has one input $i$ and a set of outputs $O$. \textit{\textbf{Pins}} and \textit{\textbf{ActivityParameterNodes}} ($APN$) are elements that hold tokens of data type $\delta$ and represent inputs and outputs of $actions$ or an $activities$, respectively. These elements can be connected by using \textit{\textbf{edges}}\footnote{Incoming and outgoing edges are treated as object flows connected between pins of type \textit{ControlToken} (CT) in order to facilitate the semantics specification.}. Requirements of the tokens transfer are defined by the $edge$ and by the target token holder ($Pin$ or $APN$). On one hand, an $edge$ allows transferring tokens that satisfy its $guard$ and only if there is at least a minimum amount of tokens ($weight$). On the other hand, a token holder has a limited space ($upperbound$), and accepts  only a specific amount of tokens per execution of the corresponding node. This amount is between the $lower$ value and the $upper$ value. $ordering$ defines the order in which tokens have to be saved in the token holder with regard to the arrival (e.g. FIFO). $streaming$ allows $APN$ to receive tokens during the execution of the activity. Using $exception$ outputs, an activity can yield an exception during the execution.

\section{Reference Semantics}

This section introduces a reference semantics of activities based on the UML-standard. This paper focuses on the semantics related to tokens transfer, hierarchy, streaming, parameter sets and control nodes.  The reference semantics is specified using SOS. This type of semantics describes changes in the system behavior by using labeled transitions ($ s \overset{l} \rightarrow s'$). In the SOS for activities, changes in the state of the model are called steps. Due to the complexity of the activity semantics, we propose two types of steps \textit{micro-steps} and  \textit{macro-steps}. While the \textit{macro-steps} involve changes on the state of nodes, \textit{micro-steps} are used to model intermediates transitions.

\subsection{State of activity execution}

The behavior of an activity is mainly based on coordinated executions of nodes, which is determined by the location of the tokens and the occurrence of events. Therefore, the state of a UML model is defined by the status of nodes ($S_n$),  activities ($S_a$), token holders ($S_{th}$), and event occurrences ($S_{\Sigma}$). This information is enough to determine all possible following behaviors of the activity in any step of the execution. 

\begin{definition}
Given the model $M = \langle A, \Sigma, D , P_{\Sigma} \rangle$, where $A$ is a set of activities, $\Sigma$ is a set of the events names, $D$ is a set of data types and $P_{\Sigma}$ is a set of pools that contains occurred events. The state of the system is given by the tuple $\langle  S_n, S_a,  S_{th}, S_{\Sigma} \rangle$, where:
\begin{itemize}
\item $S_n \in (N \rightarrow \lbrace idle \rbrace \cup \lbrace \langle executing, f_{in} \rangle | f_{in} $is a function$\rbrace)$, where  $\forall n \in N$, if $S_n(n) = \langle executing, f_{in}\rangle$ then $f_{in} \in ( input(n) \rightarrow D^* )$.

\item $S_a \in (A \rightarrow \lbrace idle \rbrace \cup \lbrace \langle executing, P_s, P_n \rangle \rbrace \cup \lbrace \langle exception(v) | v \in D \rangle\rbrace)$, where $P_s$ is the parameter set that has invoked the activity, and $P_n$ is a set of $APN$ that have to be set before the activity finishes.

\item $S_{th} \in ((P \cup APN) \rightarrow  D^*  )$, where $ \forall h \in (P \cup APN)$, if $S_{th}(h) = V$ then $ \forall v \in V.$ $ v \in \delta(h)$ and $|V|\leq upperbound(h)$.

\item $S_{\Sigma} \in ( pool \rightarrow O)$, where $O$ is the set of occurred events in the pool.
\end{itemize}
\end{definition}

\subsection{Step semantics}
SOS specifies the behavior of a system in terms of inference rules, which determine the valid transitions of the state of the system. An inference rule is applicable if its premises (above a horizontal line) are true and leads to a conclusion (below a horizontal line) that updates the state of the system. The following defines the two proposed steps types.

\begin{definition}A \textit{macro-step} ($\twoheadrightarrow$) is a transition in the state of an activity that leads to a change in the status of a node.  
\end{definition}

\begin{definition}A \textit{micro-step} ($\rightsquigarrow$) is a transition in the state of an activity that does not lead to a change in the status of any node. 
\end{definition}

\begin{figure}[!h]
\centering
\subfigure[Activity] {\includegraphics[width=0.25\textwidth]{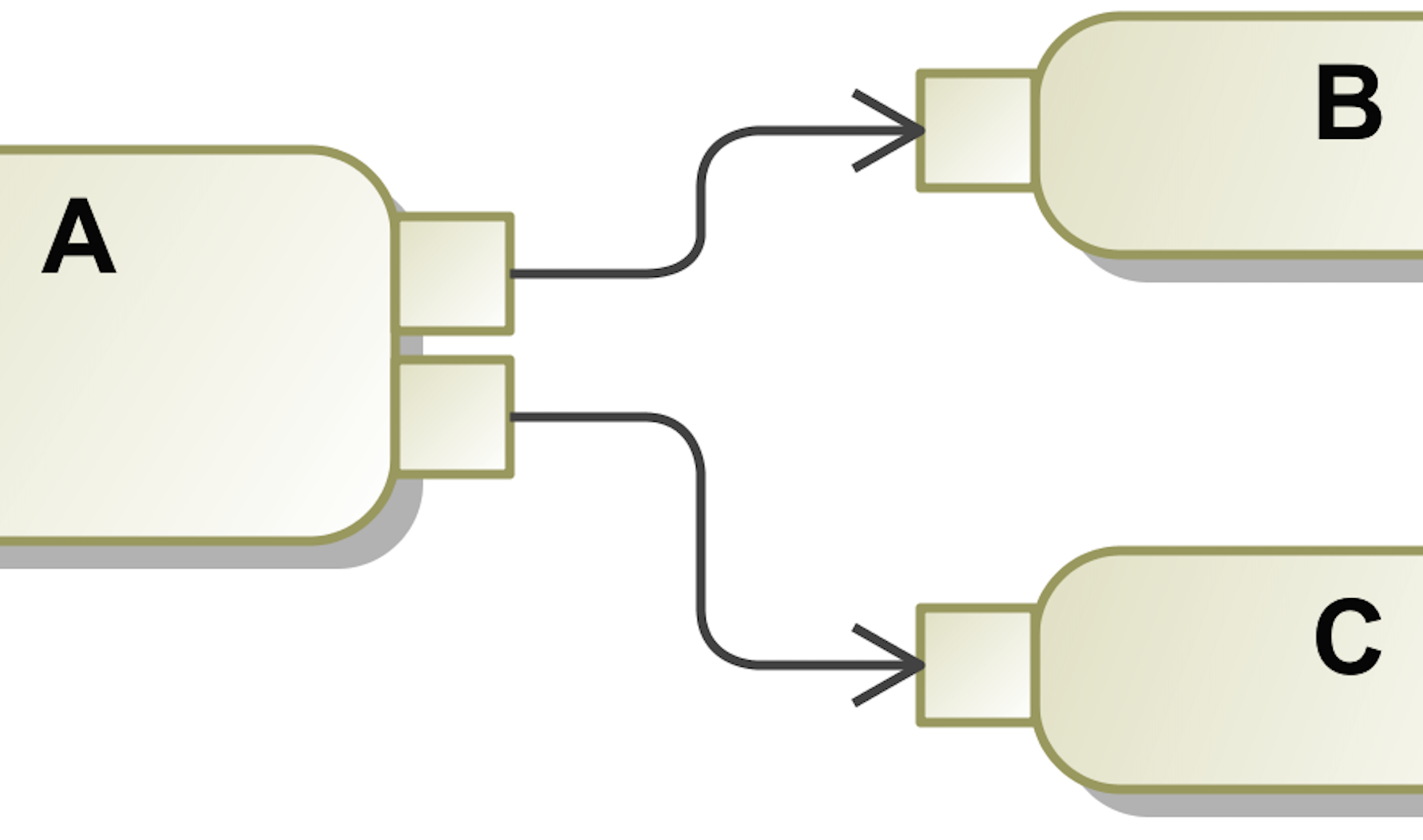} \label{fig:Act}}
\subfigure[Complete state-space] {\includegraphics[width=0.4\textwidth]{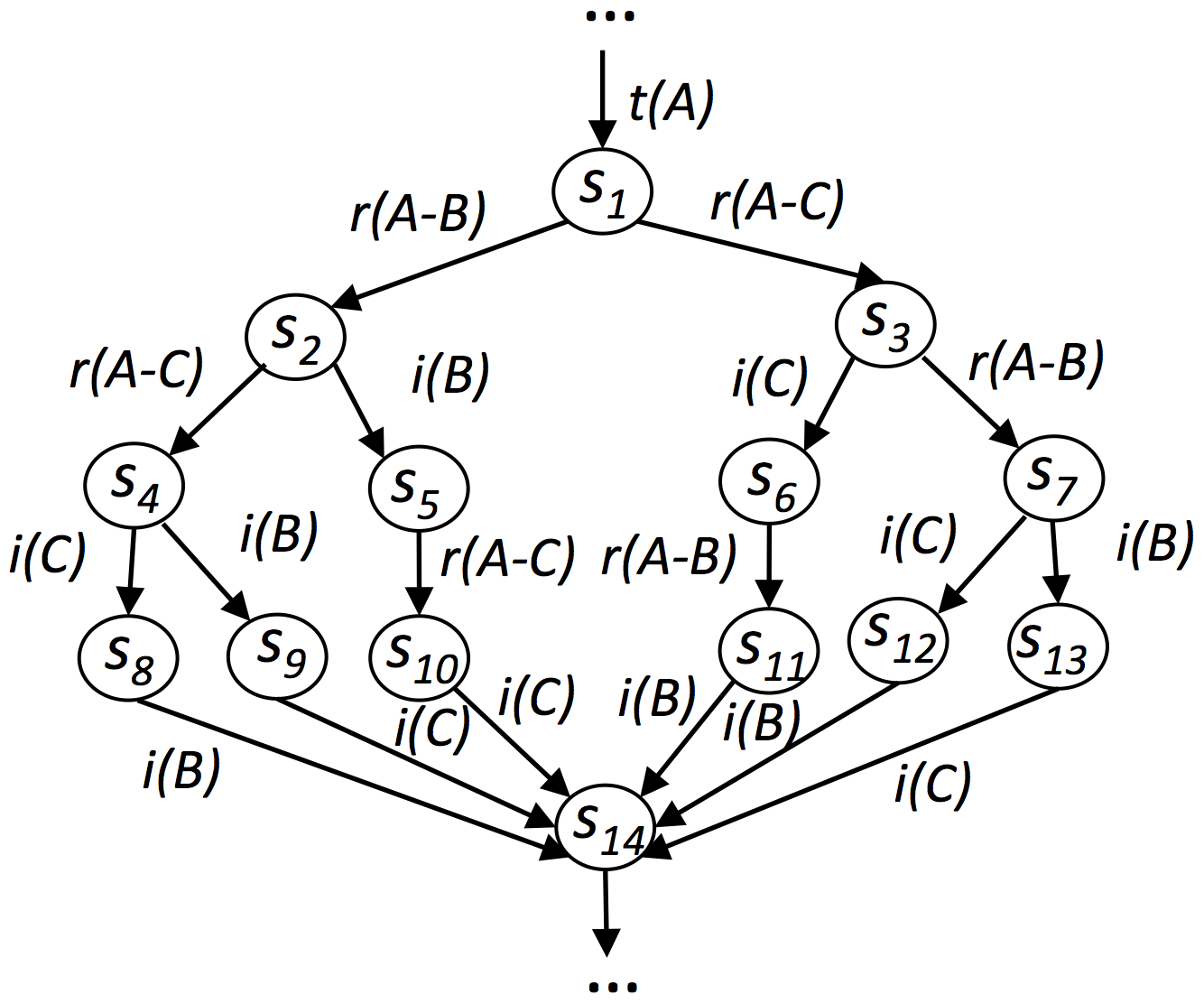} \label{fig:Comp}}
\subfigure[Reduced state-space] {\includegraphics[width=0.3\textwidth]{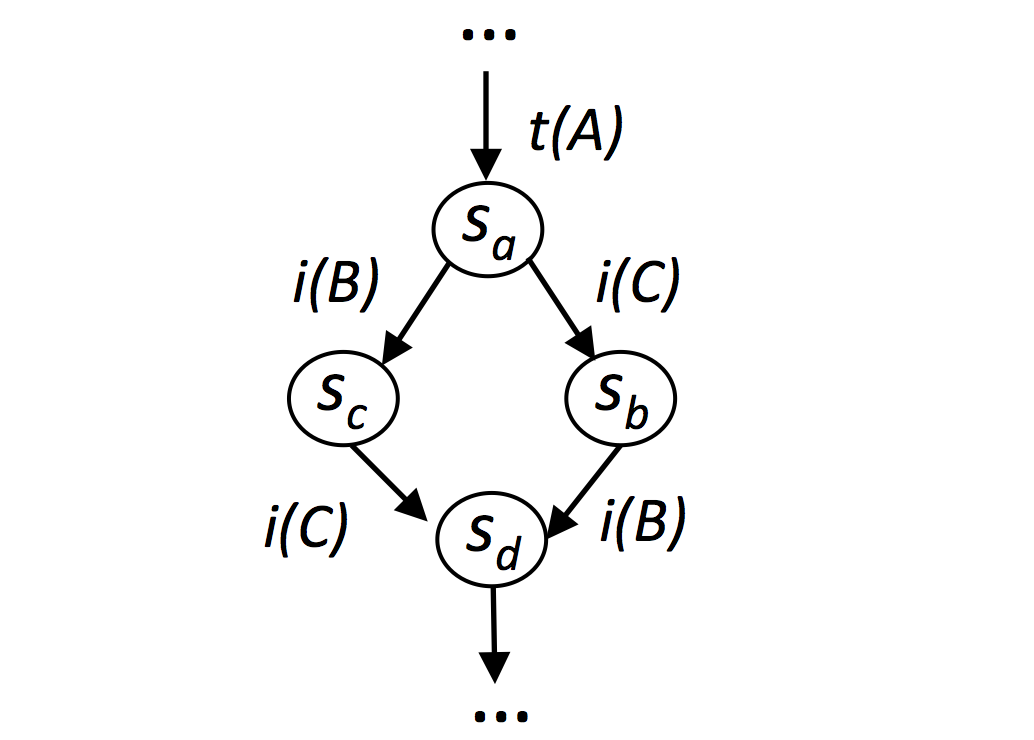} \label{fig:Red}}
\caption{Reduction of the state-space of an activity by abstracting only the information about nodes execution. Transition labels indicate termination of the execution (\textit{t}), starting of the execution (\textit{i}), and transfer of tokens (\textit{r})}
\label{fig:RedSS}
\end{figure}

This distinction is also made in order to facilitate model checking by reducing the state-space, since requirements to verify primarily refer to nodes execution (e.g. A $\square$ (received event  $\Rightarrow$  A$\lozenge$ motor starts)). Thus, the state-space can be reduced by taking only information about the execution into account.  Figure \ref{fig:Comp} shows the state-space on the execution of the actions of figure \ref{fig:Act}. Note that between the end of the execution of action \textit{A} (\textit{t(A)}) and the beginning of the execution of action \textit{B} (\textit{i(B)}) or action \textit{C} (\textit{i(C)}) there are several intermediate states, which represent tokens transfer (e.g. \textit{r(A-B)}). These intermediate states are removed in the reduced state-space (Figure \ref{fig:Red}). 

The reduced state-space is used to verify UML models and the conformity of an interpretation with the standard. In order to differentiate  transitions of both state-spaces, transitions in the reduced state-space are called \textit{transitions}. A \textit{transition} is defined as a sequence of \textit{micro-steps} (which can be empty) that ends with a \textit{macro-step} i.e., a transition ends with the start or the finalization of the execution of a node as is shown in definition 4. The \textit{transition} inherits the label \textit{l} of the \textit{macro-step}.

\newpage
\vspace{-0.5cm}

\textit{Transitions} have additional requirements (Lines 2-4) related to a subset of control nodes. In the presented work, this sub set is called switch nodes (SN) and is composed of the node types: \textit{Join}, \textit{Fork}, \textit{Merge}, and \textit{Decision}. Since SN cannot hold any token \cite[p. 327]{UML241}, the token flow through any SN has to end in the execution of a node not belonging to this subset. This condition is evaluated at the target state of the \textit{macro-step}. $N$ is a set of all nodes in the model.

\begin{definition}
A \textit{transition} ($\rightarrow$) is defined 
as follow:
\vspace{-1cm}
\begin{table}[h]
\begin{minipage}{0.98\linewidth}\centering\footnotesize
\begin{equation*}
 \frac{\parbox{4.3in} {\centering
 $ \langle S_n, S_a,  S_{th}, S_{\Sigma}  \rangle ( \rightsquigarrow)^*  \langle S_n', S_a', S_{th}', S_{\Sigma}'  \rangle , \langle S_n', S_a', S_{th}', S_{\Sigma}' \rangle \overset{l} \twoheadrightarrow \langle S_n'', S_a'', S_{th}'', S_{\Sigma}'' \rangle, $ 
 $\forall x \in N, type(x)\in SN, \forall p \in inpin(x),S_{th}''(p)=\emptyset, $\\
$ \forall y \in N,type(y)\in SN -\lbrace Fork\rbrace ,\forall q \in outpin(y), S_{th}''(q)=\emptyset,$\\
$ \forall z \in N , type(z) = Fork, \exists p \in outpin(z), S_{th}''(p)=\emptyset$
}}{ \parbox{4.3in} {\centering
$\langle S_n, S_a,  S_{th}, S_{\Sigma} \rangle  \overset{l} \rightarrow \langle S_n'', S_a'', S_{th}'', S_{\Sigma}'' \rangle$
 }}
\end{equation*}
\end{minipage}
\begin{minipage}{0.01\linewidth}\scriptsize \begin{flushleft}
\vspace{0.4cm} {\ttfamily \bfseries 1}\\
\vspace{0.15cm}	{\ttfamily \bfseries 2}\\
\vspace{0.15cm}	{\ttfamily \bfseries 3}\\
\vspace{0.15cm}	{\ttfamily \bfseries 4}\\
\vspace{0.3cm}{\ttfamily \bfseries 5}
\end{flushleft}\end{minipage}
\end{table}
\end{definition}
\vspace{-0.8cm}

A sequence of SN represents a challenge to the formal specification because it has to be first analyzed whether at the end of the token flow at least one non-SN can be executed and post conditions of the SN are satisfied (Lines 2-4). Therefore, the execution of SN is defined by a \textit{micro-step} in order to analyze all possible token flows of a sequence of SN without changing the state of the activity. This is possible because a \textit{transition} can only end with a \textit{macro-step} and, therefore, no \textit{transition} is created for token flows that do not end in an execution of a non-SN and or do not satisfy constraints for SN.

\section{Semantics of model elements}
This section presents a set of 37 inference rules that specify the behavior of model element of the activity diagrams defined in the UML standard. This rules have been implemented and tested using a simulator \cite{Daw2016}.

\subsection{Token transfer}

Tokens are transfer from a source token holder to a target token holder that are connected by an $edge$. The function \textit{transfer} defines a sequence of tokens, if any, to transfer according to the preconditions, explained in section \ref{sec:Structure}. This function ensures that the size of the sequence is bigger enough to cross the edge ($weight$) and to invocate the target node ($lower$), that the size of the target pin ($upperbound$) is not violated, all transfer tokens are immediately consumed ($upper$), and all tokens of the resulting sequence satisfy the guard.

\begin{definition}$ \text{transfer}: E \rightarrow D^*$
\footnotesize
\[ \text{transfer(e)} = \left\{
  \begin{array}{lr}
	\emptyset & \text{\textbf{if} } |V_o| <  lower(target(e)) \wedge \\
	& |V_o| < weight(e)\\
	&\\
	\lbrace v1,...,vk \rbrace = \lbrace v \in V_o | k=max \rbrace & \text{\textbf{elseif }} type(node(target(e)) \not \in SN\\
	  &\\
	\lbrace v1,...,vk \rbrace = \lbrace v \in V_o | k \leq max \wedge & \text{\textbf{else}} \\
  k \geq lower(target(e)) \wedge &\\
 k \geq weight(e)  \rbrace &\\  
  \end{array}
\right.
\] where,
\footnotesize
 $V_{o} = \lbrace v \in S_{th}(source(e)) |   guard(e)(v)=true  \rbrace $, and\\
 $max = min(|V_o|,upper(target(t)), upperbound(target(t)))$

\end{definition}

SN act as a switch in the token transfer. Therefore, the properties of the transfer are defined by the sources and the targets of the switch, which are the ones that can hold tokens. In order to analyze all possible tokens transfer, the function \textit{transfer} non-deterministically chooses a sequence size ($k$) for SN. The allowed sizes are limited by the preconditions of the token transfer as is shown in the else-statement. In addition to the above-presented preconditions, tokens can be transfer only if they can be immediately consumed by the target node \cite[p. 320]{UML241}. This implies that the consumption of all required tokens and the beginning of the node execution are performed in the same \textit{macro-step}.

\subsection{Action}

\textbf{Token consumption and invocation:}
An example of a \textit{macro-step} is shown in the following SOS-rule, which determines the invocation of an $action$. This rule is applicable only for inactive $actions$ that are contained by $activities$ that are executing (Line 1). Furthermore, all input pins must have been offered enough tokens for the execution. Note that the function \textit{transfer} determines if the tokens transfer is possible according to the preconditions of the edge, and the target pin. Since an input pin can have multiple $edges$, and thereby multiple source pins, this rule non-deterministically chooses a source  by using the function $F_{tl}$. This function returns a set of injective functions that map target pins into source pins, which offer tokens to consume.  

$Vc_i$ defines the tokens to consume (Line 3) and $V_{q_i}$ defines the tokens that are in the source pins (Line 6). Thus, the source pins are updated with the difference~($\rceil$) of these two sequences (Line 8). The values of the tokens to consume are saved in the action's state using the function  $f_{in}$ (Line 4), which maps input pins to sequences of values. This function forms part of the status of the action after the update (Line 7). This function in combination with $m_{io}$ is used in the termination of the action to define which sequence of tokens are offered in the output pins.  Note that before the mapping the tokens sequence is reordered (Line 4). After executing this rule, the termination rule of the action becames applicable.

\begin{table}[h]\begin{minipage}{0.98\linewidth}\centering \footnotesize
\begin{equation*}
\frac{\parbox{4.5in} {
$a \in A, n \in node(a),type(n) = Action,S_a(a)=\langle executing, P_s, P_n\rangle, S_n(n)= idle, $\\
$ f_{tl} \in F_{tl}(inpin(n)),\forall p \in inpin(n). transfer(f_{tl}(p)) \neq \emptyset,$\\
 $ \lbrace  Vc_1, ..., Vc_i\rbrace = \lbrace transfer(f_{tl}(p)) | \forall p \in inpin(n)\rbrace,$\\
 $ \lbrace  p_1, ..., p_i \rbrace = inpin(n), f_{in}\triangleq ordering(p_k)(Vc_k), \forall 1 \leq k \leq i,$\\
 $ \lbrace  q_1, ..., q_i \rbrace = \lbrace source(f_{tl}(p_k))| \forall 1 \leq k \leq i\rbrace,$\\
$\forall 1 \leq k \leq i.V_{q_k}=S_{th}(q_k)$
}}{ \parbox{4.5in} {
$\langle   S_n, S_a,  S_{th}, S_{\Sigma}  \rangle \overset{i(n)} \twoheadrightarrow \langle S_n [ n \mapsto \langle executing, f_{in} \rangle], S_a, $\\
$S_{th}[ q_1 \mapsto V_{q_1} \rceil  Vc_1]...[ q_i \mapsto V_{q_i} \rceil  Vc_i],  S_{\Sigma}\rangle$
 }}
\end{equation*}
\end{minipage}
\begin{minipage}{0.01\linewidth}\scriptsize \begin{flushleft}
\vspace{0.1cm} {\ttfamily \bfseries 1}\\
\vspace{0.11cm}	{\ttfamily \bfseries 2}\\
\vspace{0.11cm}	{\ttfamily \bfseries 3}\\
\vspace{0.11cm}	{\ttfamily \bfseries 4}\\
\vspace{0.11cm}	{\ttfamily \bfseries 5}\\
\vspace{0.11cm}	{\ttfamily \bfseries 6}\\
\vspace{0.3cm}{\ttfamily \bfseries 7}\\
\vspace{0.11cm}	{\ttfamily \bfseries 8}
\end{flushleft}\end{minipage}
\end{table}

\textbf{Execution Termination and token offering: } This rules is applicable once the action is been executing (Line 2).  The outputs are calculated by using the mapping function $m_{io}$ and inputs values (Line 3), which are associated to the state \textit{executing} via  the function $f_{in}$. Note that the resulting tokens have to satisfy the multiplicity of output pins (Lines 7-8).

\begin{table}[!h] \begin{minipage}{0.98\linewidth}\centering \footnotesize
\begin{equation*} \frac{\parbox{4.7in} {
 $a \in A, n \in node(a), type(n) = Action, S_a(a)=\langle executing, P_s, P_n \rangle, $\\
 $S_n(n)= \langle executing, f_{in} \rangle,$\\
 $m_{io}(n)(f_{in}) = r \in outpin(n) \rightarrow D^*, $\\
$\lbrace  q_1, ..., q_i \rbrace = outpin(n),$\\
$\forall 1 \leq k \leq i.V_{q_k}=S_{th}(q_k),$\\
$\forall 1 \leq k \leq i.Vo_{k}=ordering(q_k)(r(q_k)),$\\
$ \forall q_k \in outpin(n). | V_{q_k} \cup Vo_k |\leq upperbound(q_k) ,$\\
$ \forall q_k \in outpin(n). | Vo_k | \geq lower(q_k) $
}}{ \parbox{4.7in} {
$\langle  S_n, S_a,  S_{th}, S_{\Sigma}  \rangle \overset{t(n)} \twoheadrightarrow \langle  S_n [ n \mapsto idle], S_a, S_{th}[ q_1 \mapsto V_{q_1} \cup Vo_1]...[ q_i \mapsto V_{q_i} \cup Vo_i], S_{\Sigma} \rangle$
 }}
\end{equation*}
\end{minipage}
\begin{minipage}{0.01\linewidth}\scriptsize \begin{flushleft}
\vspace{0cm} {\ttfamily \bfseries 1}\\
\vspace{0.11cm}	{\ttfamily \bfseries 2}\\
\vspace{0.11cm}	{\ttfamily \bfseries 3}\\
\vspace{0.11cm}	{\ttfamily \bfseries 4}\\
\vspace{0.11cm}	{\ttfamily \bfseries 5}\\
\vspace{0.11cm}	{\ttfamily \bfseries 6}\\
\vspace{0.11cm}	{\ttfamily \bfseries 7}\\
\vspace{0.11cm} {\ttfamily \bfseries 8}\\
\vspace{0.33cm}	{\ttfamily \bfseries 9}
\end{flushleft}\end{minipage}
\end{table}

\subsection{InitialNode}
Initial node is invoked during the invocation of the activity to which the node belogs. Therefore, the initial node has only one inferece rule. 

\textbf{Termination and offering tokens:}
Once the node is executing, the termination rule becomes applicable. Initial nodes offers a \textit{ControlToken} in its output after finalizing the execution.

\begin{table}[!h] \begin{minipage}{0.98\linewidth}\centering \footnotesize
\begin{equation*} 
\frac{\parbox{4.7in} {
\vspace{-0.4cm}
$a \in A, n \in node(a),type(n) = InitialNode, S_a(a)=\langle executing, P_s, P_n\rangle, $\\
$S_n(n)= executing, $\\
$ q = outpin(n)$
}}{ \parbox{4.7in} {
$\langle   S_n, S_a,  S_{th}, S_{\Sigma}  \rangle \overset{i(n)} \twoheadrightarrow \langle S_n [ n \mapsto \langle idle, \emptyset \rangle], S_a, S_{th}[ q \mapsto \lbrace CT\rbrace], S_{\Sigma} \rangle$
 }}
\end{equation*}
\end{minipage}
\begin{minipage}{0.01\linewidth}\scriptsize \begin{flushleft}
\vspace{-0.25cm} {\ttfamily \bfseries 1}\\
\vspace{0.11cm}	{\ttfamily \bfseries 2}\\
\vspace{0.11cm}	{\ttfamily \bfseries 3}\\
\vspace{0.33cm}	{\ttfamily \bfseries 4}\\

\end{flushleft}\end{minipage}
\end{table}


\subsection{ForkNode}
The following SOS-rule specifies token consumption and activation of a \textit{Fork}, which is shown as an example of SN. A \textit{Fork} creates a token in each output for each incoming token. An incoming token is consumed by a \textit{Fork} only if at least one of the outputs offers is accepted, i.e. immediately consumed by the target node. Outgoing tokens that cannot immediately be consumed remain in the output except for tokens that do not satisfy the guard of the outgoing edge. This constraint is also evaluated in the definition of a transition (Definition 4, Line 4). The function \textit{transfer} defines the tokens sequence $Vo$ that can be offered by the target node (Line 3). Therefore, this rule is applicable for all possible sizes of sequences returned by this function. $Vc$ defines a sub set of the offered tokens that can satisfy the guards of at least one outgoing edge (Line 4). This subset ensures that at least one of the outputs offers is accepted. The precondition of the edge is reevaluated with the set of tokens to consume (Line 5). In case that all premises are satisfied, the state of the pins is updated (Line 11-12). A termination rule sets the sequence of tokens in the outputs (see~Appendix).

\begin{table}[h] \begin{minipage}{0.98\linewidth}\centering \footnotesize
\begin{equation*}\frac{\parbox{4.7in} {
\vspace{-0.4cm}
$a \in A, n,m \in node(a),type(n) = Fork,S_a(a)=\langle executing, P_s, P_n\rangle, S_n(n)= idle, $\\
$ e \in E. e= \langle s, t, g, w\rangle \wedge t \in inpin(n) \wedge transfer(e) \neq \emptyset,$\\
$  Vo =  transfer(e),$\\
$ Vc= \lbrace v \in Vo | \exists p \in outpin(n).  \exists e \in edge(a). e=\langle p, t', g', w'\rangle \wedge g'(v)=true\rbrace,$\\
$|Vc| \geq weight(e), $\\
  $ V_s = S_{th}(source(e)),$
}}{ \parbox{4.7in} {
$\langle S_n, S_a,  S_{th}, S_{\Sigma} \rangle \rightsquigarrow \langle S_n[\langle executing, \emptyset \rangle], S_a,  S_{th}[source(e) \mapsto V_s \rceil Vc]$\\
$[target(e) \mapsto Vc], S_{\Sigma} \rangle$
 }}
\end{equation*}
\end{minipage}
\begin{minipage}{0.01\linewidth}\scriptsize \begin{flushleft}
\vspace{-0.25cm} {\ttfamily \bfseries 1}\\
\vspace{0.11cm}	{\ttfamily \bfseries 2}\\
\vspace{0.11cm}	{\ttfamily \bfseries 3}\\
\vspace{0.11cm}	{\ttfamily \bfseries 4}\\
\vspace{0.11cm}	{\ttfamily \bfseries 5}\\
\vspace{0.11cm}	{\ttfamily \bfseries 6}\\
\vspace{0.3cm} {\ttfamily \bfseries 7}\\
\vspace{0.11cm}	{\ttfamily \bfseries 8}
\end{flushleft}\end{minipage}
\end{table}

\textbf{Termination and token offering:} Tokens that satisfy the guards are offered in the outputs (Line 6). Note that invocation rule of the \textit{Fork} (Section \ref{sec:Switch nodes}) ensures that all tokens satisfy at least the guard of one output.
\vspace{-0.8cm}
\begin{table}[!h] \begin{minipage}{0.98\linewidth}\centering \footnotesize
\begin{equation*} \frac{\parbox{4.7in} {
$a \in A, n,m \in node(a),type(n) = Fork,S_a(a)=\langle executing, P_s, P_n\rangle, $\\
$S_n(n)= \langle executing,\emptyset \rangle, $\\
$ Vo = S_{th}(inpin(n)),$\\
$\lbrace q_1,..q_i \rbrace =outpin(n),$\\
 $  \lbrace  V_{q_1}, ..., V_{q_i}\rbrace \subseteq D^*,$\\
 $\forall 1 \leq k \leq i. V_{q_k}= \lbrace v \in Vo | \exists e \in edge(a). e=\langle q_k, t', g', w'\rangle \wedge g'(v)=true\rbrace,$\\
 $\forall 1 \leq k \leq i.V_{r_k}=S_{th}(q_k),$
}}{ \parbox{4.7in} {
$\langle S_n, S_a,  S_{th}, S_{\Sigma} \rangle  \rightsquigarrow \langle S_n [n \mapsto idle], S_a,  S_{th}[inpin(n) \mapsto \emptyset]$\\
$[q_1 \mapsto V_{r_1} \cup  V_{q_1}]...[q_i \mapsto V_{r_i} \cup  V_{q_i}], S_{\Sigma} \rangle$
 }}
\end{equation*}
\end{minipage}
\begin{minipage}{0.01\linewidth}\scriptsize \begin{flushleft}
\vspace{0.25cm} {\ttfamily \bfseries 1}\\
\vspace{0.11cm}	{\ttfamily \bfseries 2}\\
\vspace{0.11cm}	{\ttfamily \bfseries 3}\\
\vspace{0.11cm}	{\ttfamily \bfseries 4}\\
\vspace{0.11cm}	{\ttfamily \bfseries 5}\\
\vspace{0.11cm}	{\ttfamily \bfseries 6}\\
\vspace{0.11cm}	{\ttfamily \bfseries 7}\\
\vspace{0.3cm} {\ttfamily \bfseries 8}\\
\vspace{0.11cm}	{\ttfamily \bfseries 9}
\end{flushleft}\end{minipage}
\end{table}

\clearpage
\newpage
\subsection{JoinNode}

\textbf{Token consumption and invocation:}
A join node is executed if the correspoding \textit{JoinSpecification} is true (Line 4). This specification determines which inputs are required to have tokens for the execution. If multiple \textit{ControlTokens} are offered in one input, the token are merged into one tokens by using the function \textit{combine}. Data tokens that are offered in different inputs are offered in the output as an ordered set. The order is given by the order of offering, which is defined in the state by $P$.

\begin{table}[!h] \begin{minipage}{0.98\linewidth}\centering \footnotesize
\begin{equation*} 
\frac{\parbox{4.7in} {
\vspace{-0.4cm}
$a \in A, n,m \in node(a),type(n) = Join,$\\
$S_a(a)=\langle executing, P_s, P_n\rangle, S_n(n)= \langle idle, P \rangle , $\\
$joinSpec(inpun(n))=true,$\\
$ f_{tl} \in F_{tl}(inpin(n)), \forall p \in inpin(n). transfer(f_{tl}(p)) \neq \emptyset,$  \\
$ \forall p \in inpin(n). \delta(p) \neq ControlToken \wedge p \in P,$\\
$\lbrace Vc_{p_1}, ..., Vc_{p_i} \rbrace = \lbrace combine(transfer(f_{tl}(p))) | \forall t \in inpin(p)\rbrace,$\\
$\lbrace p_1, ..., p_i \rbrace = inpin(n), f_{in}\triangleq index(P,p_k), \forall 1 \leq k \leq i,$\\
$\lbrace q_1, ..., q_i \rbrace =\lbrace s | \forall t \in inpin(p).\exists e= \langle s, t, g, w\rangle  \rbrace,$\\
$ \forall 1 \leq k\leq i.  Vc_{p_k}= combine(transfer(f_{tl}(p_k))),$\\
$ \forall 1 \leq k\leq i.  V_{q_k}= S_{th}(q_k)$
}}{ \parbox{4.7in} {
$\langle S_n, S_a,  S_{th}, S_{\Sigma} \rangle \overset{i(n)}  \rightsquigarrow \langle S_n[n \mapsto \langle executing, f_{in} \rangle], S_a,  S_{th}[p_1 \mapsto Vc_{p_1}]...[p_i \mapsto Vc_{p_i}]$\\
$[q_1 \mapsto V_{q_1} \rceil   Vc_{p_1}]...[q_i \mapsto V_{q_i} \rceil  Vc_{p_i}], S_{\Sigma} \rangle$
 }}
\end{equation*}
\end{minipage}
\begin{minipage}{0.01\linewidth}\scriptsize \begin{flushleft}
\vspace{-0.25cm} {\ttfamily \bfseries 1}\\
\vspace{0.11cm}	{\ttfamily \bfseries 2}\\
\vspace{0.11cm}	{\ttfamily \bfseries 3}\\
\vspace{0.11cm}	{\ttfamily \bfseries 4}\\
\vspace{0.11cm}	{\ttfamily \bfseries 5}\\
\vspace{0.11cm}	{\ttfamily \bfseries 6}\\
\vspace{0.11cm}	{\ttfamily \bfseries 7}\\
\vspace{0.11cm} {\ttfamily \bfseries 8}\\
\vspace{0.11cm}	{\ttfamily \bfseries 9}\\
\vspace{0.11cm} {\ttfamily \bfseries 10}\\
\vspace{0.11cm}	{\ttfamily \bfseries 11}\\
\vspace{0.33cm} {\ttfamily \bfseries 12}\\
\vspace{0.11cm}	{\ttfamily \bfseries 13}
\end{flushleft}\end{minipage}
\end{table}

\textbf{Saving the order of token offering - adding:}
This rules aims to save the order in which the tokens are offered by saving the pin in the state of the node.

\begin{table}[!h] \begin{minipage}{0.98\linewidth}\centering \footnotesize
\begin{equation*} 
\frac{\parbox{4.7in} {
\vspace{-0.4cm}
$a \in A, n,m \in node(a),type(n) = Join,S_a(a)=\langle executing, P_s, P_n\rangle,$\\
$ S_n(n)= \langle idle, P_{order}\rangle, $\\
$ p \in inpin(n). p \not \in P_{order} \wedge p \neq ControlToken \wedge$\\
$ \exists e \in edge(a). e= \langle s, p, g, w\rangle  \wedge transfer(e) \neq \emptyset$
}}{ \parbox{4.7in} {
$\langle S_n, S_a,  S_{th}, S_{\Sigma} \rangle   \overset{\tau} \twoheadrightarrow \langle S_n[n \mapsto \langle idle,  P_{order} \cup p \rangle], S_a,  S_{th}, S_{\Sigma} \rangle$
 }}
\end{equation*}
\end{minipage}
\begin{minipage}{0.01\linewidth}\scriptsize \begin{flushleft}
\vspace{-0.25cm} {\ttfamily \bfseries 1}\\
\vspace{0.11cm}	{\ttfamily \bfseries 2}\\
\vspace{0.11cm}	{\ttfamily \bfseries 3}\\
\vspace{0.11cm}	{\ttfamily \bfseries 4}\\
\vspace{0.33cm}	{\ttfamily \bfseries 5}
\end{flushleft}\end{minipage}
\end{table}

\textbf{Saving the order of token offering - removing:} This rule is applicable in case the a tokens that have been offered to the join node has been consumed by other node. This is necesary in order to update the list of offering pins from the state of the node. 

\begin{table}[!h] \begin{minipage}{0.98\linewidth}\centering \footnotesize
\begin{equation*} 
\frac{\parbox{4.7in} {
\vspace{-0.4cm}
$a \in A, n,m \in node(a),type(n) = Join,S_a(a)=\langle executing, P_s, P_n\rangle, $\\
$S_n(n)= \langle idle, P_{order}\rangle, $\\
$ p \in inpin(n). p \in P_{order} \wedge p \neq ControlToken \wedge$\\
$ \exists e \in edge(a). e= \langle s, p, g, w\rangle  \wedge transfer(e) = \emptyset$
}}{ \parbox{4.7in} {
$\langle S_n, S_a,  S_{th}, S_{\Sigma} \rangle \overset{\tau}  \twoheadrightarrow \langle S_n[n \mapsto \langle idle,  P_{order} \rceil p \rangle], S_a,  S_{th}, S_{\Sigma} \rangle$
 }}
\end{equation*}
\end{minipage}
\begin{minipage}{0.01\linewidth}\scriptsize \begin{flushleft}
\vspace{-0.25cm} {\ttfamily \bfseries 1}\\
\vspace{0.11cm}	{\ttfamily \bfseries 2}\\
\vspace{0.11cm}	{\ttfamily \bfseries 3}\\
\vspace{0.11cm}	{\ttfamily \bfseries 4}\\
\vspace{0.33cm}	{\ttfamily \bfseries 5}
\end{flushleft}\end{minipage}
\end{table}

\clearpage
\newpage
\subsection{MergeNode}
In contrast with other nodes, merge nodes required only one rule to transfer input tokens to outgoing outputs. 

\textbf{Invocation, and token consumption and offering}
Note that the transfer is carried out without invocating the node. 

\begin{table}[!h] \begin{minipage}{0.98\linewidth}\centering \footnotesize
\begin{equation*} 
\frac{\parbox{4.7in} {
\vspace{-0.4cm}
$a \in A, n,m \in node(a),type(n) = Merge,S_a(a)=\langle executing, P_s, P_n\rangle, S_n(n)= idle, $\\
$ e \in E. e= \langle s, t, g, w\rangle \wedge t \in inpin(n) \wedge transfer(e) \neq \emptyset,$\\
$Vo =  transfer(e),$\\
$p=source(e),q \in outpin(n),$\\
$V_q = \lbrace V |  S_{th}(q)=V \rbrace,$\\
$V_p = \lbrace V |  S_{th}(p)=V \rbrace,$
}}{ \parbox{4.7in} {
$\langle S_n, S_a,  S_{th}, S_{\Sigma} \rangle \overset{i(n)}  \rightsquigarrow \langle S_n, S_a,  S_{th}[q \mapsto V_{q} \cup  Vo][p \mapsto V_p \rceil Vc], S_{\Sigma} \rangle$
 }}
\end{equation*}
\end{minipage}
\begin{minipage}{0.01\linewidth}\scriptsize \begin{flushleft}
\vspace{-0.25cm} {\ttfamily \bfseries 1}\\
\vspace{0.11cm}	{\ttfamily \bfseries 2}\\
\vspace{0.11cm}	{\ttfamily \bfseries 3}\\
\vspace{0.11cm}	{\ttfamily \bfseries 4}\\
\vspace{0.11cm}	{\ttfamily \bfseries 5}\\
\vspace{0.11cm}	{\ttfamily \bfseries 6}\\
\vspace{0.33cm}	{\ttfamily \bfseries 7}
\end{flushleft}\end{minipage}
\end{table}

\vspace{-1cm}
\subsection{DecisionNode}
The entire behavior of a decision node is specified in 7 inference rules. This is because the value that is evaluated in the decision can come from the input of the node, a special input or from the result of an associated behavior.

\textbf{Token consumption and invocation:}
This rule invokates the decision node.

\begin{table}[!h] \begin{minipage}{0.98\linewidth}\centering \footnotesize
\begin{equation*} 
\frac{\parbox{4.7in} {
\vspace{-0.4cm}
$a \in A, n,m \in node(a),type(n) = Decision,$\\
$S_a(a)=\langle executing, P_s, P_n\rangle, S_n(n)= idle, $\\
$ f_{tl} \in F_{tl}(inpin(n)),\forall p \in inpin(n). transfer(f_{tl}(p)) \neq \emptyset,$\\
 $ \lbrace  p_1, ..., p_i \rbrace = inpin(n),$\\
 $ \lbrace  q_1, ..., q_i \rbrace = \lbrace source(f_{tl}(p_k))| \forall 1 \leq k \leq 2\rbrace,$\\
 $ \forall 1 \leq k \leq i. Vc_k = transfer(f_{tl}(p_k)),$\\
 $ \forall 1 \leq k \leq i. V_{q_k} = S_{th}(q_k),$\\
$  \forall 1 \leq k \leq i. |Vc_1| = |Vc_k|$
}}{ \parbox{4.7in} {
$\langle S_n, S_a,  S_{th}, S_{\Sigma} \rangle \overset{i(n)}  \rightsquigarrow \langle S_n[ n \mapsto\langle executing, \emptyset \rangle], S_a,  S_{th}[ p_1 \mapsto Vc_1]...[ p_i \mapsto Vc_i]$\\
$[ q_1 \mapsto V_{q_1} \rceil  Vc_1]...[ q_i \mapsto V_{q_i} \rceil  Vc_i], S_{\Sigma} \rangle$
 }}
\end{equation*}
\end{minipage}
\begin{minipage}{0.01\linewidth}\scriptsize \begin{flushleft}
\vspace{-0.25cm} {\ttfamily \bfseries 1}\\
\vspace{0.11cm}	{\ttfamily \bfseries 2}\\
\vspace{0.11cm}	{\ttfamily \bfseries 3}\\
\vspace{0.11cm}	{\ttfamily \bfseries 4}\\
\vspace{0.11cm}	{\ttfamily \bfseries 5}\\
\vspace{0.11cm}	{\ttfamily \bfseries 6}\\
\vspace{0.11cm}	{\ttfamily \bfseries 7}\\
\vspace{0.11cm} {\ttfamily \bfseries 8}\\
\vspace{0.33cm}	{\ttfamily \bfseries 9}\\
\vspace{0.11cm}	{\ttfamily \bfseries 10}
\end{flushleft}\end{minipage}
\end{table}

\textbf{Termination:}
This rules terminates the execution of the decision node. Note that this rule is applicable only when only when there is not token in its input. This token is consumed after the guard evaluation is executed.

\begin{table}[!h] \begin{minipage}{0.98\linewidth}\centering \footnotesize
\begin{equation*} 
\frac{\parbox{4.7in} {
\vspace{-0.4cm}
$a \in A, n,m \in node(a),type(n) = Decision,$\\
$S_a(a)=\langle executing, P_s, P_n\rangle, S_n(n)= \langle executing, \emptyset \rangle, $\\
$S_{th}(inpin(n))=\emptyset, S_a(dBehavior (n))=idle $
}}{ \parbox{4.7in} {
$\langle S_n, S_a,  S_{th}, S_{\Sigma} \rangle \overset{t(n)}  \rightsquigarrow \langle S_n[n \mapsto idle], S_a, S_{th}, S_{\Sigma} \rangle$
 }}
\end{equation*}
\end{minipage}
\begin{minipage}{0.01\linewidth}\scriptsize \begin{flushleft}
\vspace{-0.25cm} {\ttfamily \bfseries 1}\\
\vspace{0.11cm}	{\ttfamily \bfseries 2}\\
\vspace{0.11cm}	{\ttfamily \bfseries 3}\\
\vspace{0.33cm}	{\ttfamily \bfseries 4}
\end{flushleft}\end{minipage}
\end{table}

\clearpage
\newpage

\textbf{Guard evaluation over the input values and token offering:}
This micro step removes the token from the input pin and offers it to the edge, whose guards evaluate true for the input value.

\begin{table}[!h] \begin{minipage}{0.98\linewidth}\centering \footnotesize
\begin{equation*} 
\frac{\parbox{4.7in} {
\vspace{-0.4cm}
$a \in A, n,m \in node(a),type(n) = Decision,$\\
$S_a(a)=\langle executing, P_s, P_n\rangle, S_n(n)= \langle executing, \emptyset \rangle, $\\
$dFlow  = \emptyset, dBehavior  = \emptyset, S_{th}(inpin(n))\neq \emptyset,$\\
$ V_p = \lbrace v_{p_1}, ..., v_{p_i}\rbrace =  S_{th}(inpin(n))=V ,$\\
$ r \in outpin(n). \exists e \in E. e= \langle r, t, g, w\rangle \wedge g(v_{p_1})=true,$\\
$ V_r = S_{th}(r)$
}}{ \parbox{4.7in} {
$\langle S_n, S_a,  S_{th}, S_{\Sigma} \rangle \overset{\tau}  \rightsquigarrow \langle S_n, S_a, S_{th}[inpin(n) \mapsto V_p \rceil \lbrace v_{p_1} \rbrace ][r \mapsto V_r \cup \lbrace v_{p_1} \rbrace], S_{\Sigma} \rangle$
 }}
\end{equation*}
\end{minipage}
\begin{minipage}{0.01\linewidth}\scriptsize \begin{flushleft}
\vspace{-0.25cm} {\ttfamily \bfseries 1}\\
\vspace{0.11cm}	{\ttfamily \bfseries 2}\\
\vspace{0.11cm}	{\ttfamily \bfseries 3}\\
\vspace{0.11cm}	{\ttfamily \bfseries 4}\\
\vspace{0.11cm}	{\ttfamily \bfseries 5}\\
\vspace{0.11cm}	{\ttfamily \bfseries 6}\\
\vspace{0.33cm}	{\ttfamily \bfseries 7}
\end{flushleft}\end{minipage}
\end{table}

\textbf{Guard evaluation over \textit{dFlow} and token offering:}
In contrast to previous rule, this rule evaluates the guard over the value of the pin \textit{dFlow}

\begin{table}[!h] \begin{minipage}{0.98\linewidth}\centering \footnotesize
\begin{equation*} 
\frac{\parbox{4.7in} {
\vspace{-0.4cm}
$a \in A, n,m \in node(a),type(n) = Decision,$\\
$S_a(a)=\langle executing, P_s, P_n\rangle, S_n(n)= \langle executing, \emptyset \rangle, $\\
$dFlow  \neq \emptyset, dBehavior  = \emptyset,S_{th}(inpin(n))\neq \emptyset, $\\
$q=dFlow (n), p = inpin(n)-q,$\\
$V_p = \lbrace  v_{p_1}, ..., v_{p_i}\rbrace = S_{th}(p),$\\
$V_q = \lbrace  v_{q_1}, ..., v_{q_i}\rbrace = S_{th}(q) ,$\\
$ r \in outpin(n). \exists e \in E. e= \langle r, t, g, w\rangle \wedge g(v_{q_1})=true,$\\
$ V_r =  S_{th}(r)$
}}{ \parbox{4.7in} {
$\langle S_n, S_a,  S_{th}, S_{\Sigma} \rangle \overset{\tau}  \rightsquigarrow \langle S_n, S_a, S_{th}[inpin(n) \mapsto V_p \rceil \lbrace v_{p_1} \rbrace][r \mapsto V_r \cup \lbrace v_{p_1} \rbrace]$\\
$[q \mapsto V_q \rceil \lbrace v_{q_1} \rbrace], S_{\Sigma} \rangle$
 }}
\end{equation*}
\end{minipage}
\begin{minipage}{0.01\linewidth}\scriptsize \begin{flushleft}
\vspace{-0.25cm} {\ttfamily \bfseries 1}\\
\vspace{0.11cm}	{\ttfamily \bfseries 2}\\
\vspace{0.11cm}	{\ttfamily \bfseries 3}\\
\vspace{0.11cm}	{\ttfamily \bfseries 4}\\
\vspace{0.11cm}	{\ttfamily \bfseries 5}\\
\vspace{0.11cm}	{\ttfamily \bfseries 6}\\
\vspace{0.11cm}	{\ttfamily \bfseries 7}\\
\vspace{0.11cm} {\ttfamily \bfseries 8}\\
\vspace{0.33cm}	{\ttfamily \bfseries 9}\\
\vspace{0.11cm}	{\ttfamily \bfseries 10}
\end{flushleft}\end{minipage}
\end{table}

\textbf{Guard evaluation over the result of \textit{dBehavior}and token offering:}
In this rule, the guard is evaluted over the value that retursn the behavior \textit{dBehavior}.

\begin{table}[!h] \begin{minipage}{0.98\linewidth}\centering \footnotesize
\begin{equation*} 
\frac{\parbox{4.7in} {
\vspace{-0.4cm}
$a \in A, n,m \in node(a),type(n) = Decision,S_a(a)=\langle executing, P_s, P_n\rangle, $\\
$S_n(n)= \langle executing, \emptyset \rangle,  dBehavior  \neq \emptyset, S_{th}(inpin(n))\neq \emptyset,$\\
$ V_p = \lbrace v_{p_1}, ..., v_{p_i}\rbrace =  S_{th}(outpin(dBehavior))=V ,$\\
$ r \in outpin(n). \exists e \in E. e= \langle r, t, g, w\rangle \wedge g(v_{p_1})=true,$\\
$ V_r = S_{th}(r)$
}}{ \parbox{4.7in} {
$\langle S_n, S_a,  S_{th}, S_{\Sigma} \rangle \overset{\tau}  \rightsquigarrow \langle S_n, S_a, S_{th}[inpin(n) \mapsto V_p \rceil \lbrace v_{p_1} \rbrace ][r \mapsto V_r \cup \lbrace v_{p_1} \rbrace], S_{\Sigma} \rangle$
 }}
\end{equation*}
\end{minipage}
\begin{minipage}{0.01\linewidth}\scriptsize \begin{flushleft}
\vspace{-0.25cm} {\ttfamily \bfseries 1}\\
\vspace{0.11cm}	{\ttfamily \bfseries 2}\\
\vspace{0.11cm}	{\ttfamily \bfseries 3}\\
\vspace{0.11cm}	{\ttfamily \bfseries 4}\\
\vspace{0.11cm}	{\ttfamily \bfseries 5}\\
\vspace{0.33cm}	{\ttfamily \bfseries 6}
\end{flushleft}\end{minipage}
\end{table}

\clearpage
\newpage

\textbf{Invoking decision behavior:}
In case that the decision node has a \textit{dBehavior}, the behavior is invokated by tranfering the input value of the decision node to the behavior. 

\begin{table}[!h] \begin{minipage}{0.98\linewidth}\centering \footnotesize
\begin{equation*} 
\frac{\parbox{4.7in} {
\vspace{-0.4cm}
$a \in A, n,m \in node(a),type(n) = Decision,S_a(a)=\langle executing, P_s, P_n\rangle,$\\
$ S_n(n)= \langle executing, \emptyset \rangle, dBehavior  \neq \emptyset,$\\
$ S_a(dBehavior (n))=idle, S_{th}(inpin(n))\neq \emptyset, S_{th}(outpin(dBehavior)) = \emptyset $\\
$\forall p \in inpin(dBehavior (n)). S_{th}(p)=\emptyset,$\\
$ \lbrace  p_1, ..., p_i\rbrace =  inpin(n),$ \\
$ \lbrace  V_{p_1}, ..., V_{p_i}\rbrace = V_p = \lbrace V | S_{th}(p)=V \rbrace,$\\
$ \forall 1 \leq k \leq i. \lbrace  v_{p_{k_1}}, ..., v_{p_{k_j}}\rbrace = V_{p_k}$
}}{ \parbox{4.7in} {
$\langle S_n, S_a,  S_{th}, S_{\Sigma} \rangle \overset{\tau}  \rightsquigarrow \langle S_n, S_a, S_{th}[m_p(p_1) \mapsto  \lbrace v_{p_{1_1}} \rbrace ][m_p(p_i) \mapsto \lbrace v_{p_{i_1}} \rbrace], S_{\Sigma} \rangle$
 }}
\end{equation*}
\end{minipage}
\begin{minipage}{0.01\linewidth}\scriptsize \begin{flushleft}
\vspace{-0.25cm} {\ttfamily \bfseries 1}\\
\vspace{0.11cm}	{\ttfamily \bfseries 2}\\
\vspace{0.11cm}	{\ttfamily \bfseries 3}\\
\vspace{0.11cm}	{\ttfamily \bfseries 4}\\
\vspace{0.11cm}	{\ttfamily \bfseries 5}\\
\vspace{0.11cm}	{\ttfamily \bfseries 6}\\
\vspace{0.11cm}	{\ttfamily \bfseries 7}\\
\vspace{0.33cm} {\ttfamily \bfseries 8}
\end{flushleft}\end{minipage}
\end{table}

\textbf{Termination and token offering for dBehavior :}
After the termination of the \textit{dBehavior}, the result is tranfered to the pin that is used to evaluate the guard.

\begin{table}[!h] \begin{minipage}{0.98\linewidth}\centering \footnotesize
\begin{equation*} 
\frac{\parbox{4.8in} {
\vspace{-0.4cm}
$a,b\in A,n\in node(a),type(n)=Decision,b=dBehavior (n),$\\ 
 $S_a(a)=\langle executing, P_s', P_n' \rangle, S_n(n)= \langle executing, \emptyset\rangle, S_a(b)=\langle executing, P_s, \emptyset \rangle,$\\
$\forall th \in (holders(b)-output(b)) .S_{th}(p)=\emptyset,$\\
$\forall m \in node(b).S_n(m)=idle ,$\\
$ o = output(b)$\\
$ V_o = \lbrace V | S_{th}(o)=V \rbrace$\\
$ r \in outpin(n). \exists e \in E. e= \langle r, t, g, w\rangle \wedge \forall v \in V_o. g(v)=true,$\\
$ V_r = \lbrace V | S_{th}(r)=V \rbrace$\\
$ \lbrace  p_1, ..., p_i\rbrace =  inpin(n),$ \\
$ \lbrace  V_{p_1}, ..., V_{p_i}\rbrace = V_p = \lbrace V | S_{th}(p)=V \rbrace,$\\
$ \forall 1 \leq k \leq i. \lbrace  v_{p_{k_1}}, ..., v_{p_{k_j}}\rbrace = V_{p_k}$\\
$ \lbrace  v_{q_1}, ..., v_{q_i}\rbrace = V_q = \lbrace v \in V | S_{th}(q)=V \wedge q = inpin(n) -  dFlow (n) \rbrace,$
}}{ \parbox{4.8in} {
$\langle  S_n, S_a,  S_{th}, S_{\Sigma}  \rangle \overset{t(n)} \twoheadrightarrow \langle  S_n [ n \mapsto idle], S_a[ b \mapsto idle], S_{th}[ m_p(q_1) \mapsto V_{q_1} \cup V_{p_1}]...$\\
$[ m_p(p_i) \mapsto V_{p_i} \cup V_{p_i}][ m_p(r_1) \mapsto V_{r_1} \cup \lbrace null \rbrace]...$\\
$[ m_p(r_j) \mapsto V_{r_j} \cup \lbrace null \rbrace ][p_1 \mapsto \emptyset]...[p_i \mapsto \emptyset],
 S_{\Sigma} \rangle$

}}
\end{equation*}
\end{minipage}
\begin{minipage}{0.01\linewidth}\scriptsize \begin{flushleft}
\vspace{-0.25cm} {\ttfamily \bfseries 1}\\
\vspace{0.11cm}	{\ttfamily \bfseries 2}\\
\vspace{0.11cm}	{\ttfamily \bfseries 3}\\
\vspace{0.11cm}	{\ttfamily \bfseries 4}\\
\vspace{0.11cm}	{\ttfamily \bfseries 5}\\
\vspace{0.11cm}	{\ttfamily \bfseries 6}\\
\vspace{0.11cm}	{\ttfamily \bfseries 7}\\
\vspace{0.11cm} {\ttfamily \bfseries 8}\\
\vspace{0.11cm}	{\ttfamily \bfseries 9}\\
\vspace{0.11cm}	{\ttfamily \bfseries 10}\\
\vspace{0.11cm}	{\ttfamily \bfseries 11}\\
\vspace{0.11cm}	{\ttfamily \bfseries 12}\\
\vspace{0.33cm}	{\ttfamily \bfseries 13}\\
\vspace{0.11cm}	{\ttfamily \bfseries 14}\\
\vspace{0.11cm}	{\ttfamily \bfseries 15}
\end{flushleft}\end{minipage}
\end{table}

\subsection{FlowFinalNode}
This node does not have any outputs and consume all input tokens. 

\clearpage
\newpage

\textbf{Token consumption and invocation:}
\begin{table}[!h] \begin{minipage}{0.98\linewidth}\centering \footnotesize
\begin{equation*} 
\frac{\parbox{4.7in} {
\vspace{-0.4cm}
 $a \in A, n \in node(a),type(n) = FlowFinalNode,$\\
 $S_a(a)=\langle executing, P_s, P_n\rangle, S_n(n)= idle, $\\
$ e \in E. e= \langle s, t, g, w\rangle \wedge t \in inpin(n) \wedge transfer(e) \neq \emptyset,$\\
$Vc=transfer(e)$,\\
$V_s = \lbrace V | S_{th}(source(e))=V\rbrace$
}}{ \parbox{4.7in} {
$\langle S_n, S_a,  S_{th}, S_{\Sigma} \rangle \overset{i(n)}  \rightarrow \langle S_n[n \mapsto \langle executing, \emptyset \rangle ], S_a, 
[source(e) \mapsto V_s \rceil Vc], S_{\Sigma} \rangle$
 }}
\end{equation*}
\end{minipage}
\begin{minipage}{0.01\linewidth}\scriptsize \begin{flushleft}
\vspace{-0.25cm} {\ttfamily \bfseries 1}\\
\vspace{0.11cm}	{\ttfamily \bfseries 2}\\
\vspace{0.11cm}	{\ttfamily \bfseries 3}\\
\vspace{0.11cm}	{\ttfamily \bfseries 4}\\
\vspace{0.11cm}	{\ttfamily \bfseries 5}\\
\vspace{0.33cm}	{\ttfamily \bfseries 6}
\end{flushleft}\end{minipage}
\end{table}

\textbf{Execution termination:}
\begin{table}[!h] \begin{minipage}{0.98\linewidth}\centering \footnotesize
\begin{equation*} 
\frac{\parbox{4.7in} {
\vspace{-0.4cm}
$a \in A, n \in node(a),type(n) = FlowFinalNode,S_a(a)=\langle executing, P_s, P_n\rangle,$\\
$ S_n(n)= \langle executing, \emptyset \rangle, $
}}{ \parbox{4.7in} {
$\langle  S_n, S_a,  S_{th}, S_{\Sigma}\rangle \overset{t(n)}  \rightarrow \langle  S_n [ n \mapsto idle], S_a, S_{th}, S_{\Sigma}\rangle$
 }}
\end{equation*}
\end{minipage}
\begin{minipage}{0.01\linewidth}\scriptsize \begin{flushleft}
\vspace{-0.25cm} {\ttfamily \bfseries 1}\\
\vspace{0.11cm}	{\ttfamily \bfseries 2}\\
\vspace{0.33cm}	{\ttfamily \bfseries 3}
\end{flushleft}\end{minipage}
\end{table}

\subsection{ActivityFinalNode}
This node terminates all processing of the activity in its invokation

\textbf{Token consumption and invocation for asynchronous call:}

\begin{table}[!h] \begin{minipage}{0.98\linewidth}\centering \footnotesize
\begin{equation*} 
\frac{\parbox{4.7in} {
\vspace{-0.4cm}
 $a \in A, n,m \in node(a),type(n) = ActivityFinalNode,$\\
 $S_a(a)=\langle executing, P_s, P_n\rangle, S_n(n)= idle,
 $\\
 $type(m)=CallBehaviorAction,  a = behavior(m), isSynchronous(m)=false, $\\
$ e \in E. e= \langle s, t, g, w\rangle \wedge t \in inpin(n) \wedge transfer(e) \neq \emptyset,$\\
$ B_{async} = \lbrace b \in A | \exists n \in node(a) \wedge  b=behavior(n) \wedge isSynchronous(n)=false \rbrace,$\\
 $\lbrace 	n_1, ..., n_i \rbrace = node(a) - \lbrace n \in node(b)| b \in B_{async} \rbrace,$\\
 $\lbrace 	a_1, ..., a_j \rbrace = {a} \cup \lbrace b \in A | b \neq a \wedge \exists n \in node(a) \wedge  b=behavior(n) \wedge $\\
$ isSynchronous(n)=true\rbrace,
 $\\
  $ \lbrace  p_1, ..., p_k \rbrace = pin(a) - \lbrace p \in pin(b)| b \in B_{async} \rbrace $
  $q$
}}{ \parbox{4.7in} {
$\langle S_n, S_a,  S_{th}, S_{\Sigma} \rangle \overset{i(n)}  \rightarrow \langle S_n[ n_1 \mapsto idle]...[ n_i \mapsto idle], S_a[ a_1 \mapsto idle]...[ a_l \mapsto idle], $\\
$S_{th}[ p_1 \mapsto \emptyset]...[ p_k \mapsto \emptyset], S_{\Sigma} \rangle$
 }}
\end{equation*}
\end{minipage}
\begin{minipage}{0.01\linewidth}\scriptsize \begin{flushleft}
\vspace{-0.25cm} {\ttfamily \bfseries 1}\\
\vspace{0.11cm}	{\ttfamily \bfseries 2}\\
\vspace{0.11cm}	{\ttfamily \bfseries 3}\\
\vspace{0.11cm}	{\ttfamily \bfseries 4}\\
\vspace{0.11cm}	{\ttfamily \bfseries 5}\\
\vspace{0.11cm}	{\ttfamily \bfseries 6}\\
\vspace{0.11cm}	{\ttfamily \bfseries 7}\\
\vspace{0.11cm} {\ttfamily \bfseries 8}\\
\vspace{0.11cm}	{\ttfamily \bfseries 9}\\
\vspace{0.33cm} {\ttfamily \bfseries 10}\\
\vspace{0.11cm}	{\ttfamily \bfseries 11}
\end{flushleft}\end{minipage}
\end{table}

\clearpage
\newpage

\textbf{Token consumption and invocation for synchronous call:}
\begin{table}[!h] \begin{minipage}{0.98\linewidth}\centering \footnotesize
\begin{equation*} 
\frac{\parbox{4.7in} {
\vspace{-0.4cm}
 $a \in A, n,m \in node(a),type(n) = ActivityFinalNode,$\\
 $S_a(a)=\langle executing, P_s, P_n\rangle, S_n(n)= idle,
 $\\
 $type(m)=CallBehaviorAction,  a = behavior(m), isSynchronous(m)=true, $\\
$ e \in E. e= \langle s, t, g, w\rangle \wedge t \in inpin(n) \wedge transfer(e) \neq \emptyset,$\\
$ B_{async} = \lbrace b \in A | \exists n \in node(a) \wedge  b=behavior(n) \wedge  isSynchronous(n)=false \rbrace,$\\
 $\lbrace 	n_1, ..., n_i \rbrace = node(a) - \lbrace n \in node(b)| b \in B_{async} \rbrace,$\\
 $\lbrace 	a_1, ..., a_j \rbrace = {a} \cup \lbrace b \in A | b \neq a \wedge \exists n \in node(a) \wedge  b=behavior(n) \wedge $\\
 $ isSynchronous(n)=true\rbrace,
 $\\
  $ \lbrace  t_1, ..., t_k \rbrace = pin(a) - \lbrace p \in pin(b)| b \in B_{async} \rbrace, $\\
 $ P=\lbrace p \in outpin(a) | S_{th}=V \wedge V \neq \emptyset \rbrace, P' = outpin(a)-P,$\\
$ \lbrace  q_1, ..., q_i \rbrace = \lbrace p \in inpin(m) | m_p^{-1} (p) \in P \rbrace, $\\
 $\lbrace  r_1, ..., r_j \rbrace = \lbrace p \in inpin(m) | m_p^{-1} (p) \not \in P \rbrace, $\\
$ \lbrace V_{p_1}, ..., V_{p_i} \rbrace = \lbrace V | \forall p \in P.S_{th}(p)=V\rbrace, $\\
 $ \lbrace V_{q_1}, ..., V_{q_i} \rbrace = \lbrace  V | \forall q \in P.S_{th}(m_p^{-1}(q))=V\rbrace, $\\
$ \lbrace V_{r_1}, ..., V_{r_j} \rbrace = \lbrace  V | \forall r \in P'.S_{th}(m_p^{-1}(r))=V\rbrace, $\\
$ \forall q_k \in m_p^{-1}(P). | V_{q_k} \cup V_{p_k} |\leq upperbound(q_k) ,$\\
$ \forall r_k \in m_p^{-1}(P'). |V_{r_k}| + 1\leq upperbound(r_k) ,$
}}{ \parbox{4.7in} {
$\langle S_n, S_a,  S_{th}, S_{\Sigma} \rangle \overset{i(n)}  \rightarrow \langle S_n[ n_1 \mapsto idle]...[ n_i \mapsto idle][ m \mapsto idle],$\\
 $S_a[ a_1 \mapsto idle]...[ a_l \mapsto idle], S_{th}[ t_1 \mapsto \emptyset]...[ t_k \mapsto \emptyset][ q_i \mapsto V_{q_i} \cup V_{p_i}]$\\
 $[r_1 \mapsto V_{r_1} \cup \lbrace null \rbrace]...[ r_j \mapsto V_{r_j} \cup \lbrace null \rbrace ], S_{\Sigma} \rangle$
 }}
\end{equation*}\end{minipage}
\begin{minipage}{0.01\linewidth}\scriptsize \begin{flushleft}
\vspace{-0.25cm} {\ttfamily \bfseries 1}\\
\vspace{0.11cm}	{\ttfamily \bfseries 2}\\
\vspace{0.11cm}	{\ttfamily \bfseries 3}\\
\vspace{0.11cm}	{\ttfamily \bfseries 4}\\
\vspace{0.11cm}	{\ttfamily \bfseries 5}\\
\vspace{0.11cm}	{\ttfamily \bfseries 6}\\
\vspace{0.11cm}	{\ttfamily \bfseries 7}\\
\vspace{0.11cm} {\ttfamily \bfseries 8}\\
\vspace{0.11cm}	{\ttfamily \bfseries 9}\\
\vspace{0.11cm}	{\ttfamily \bfseries 10}\\
\vspace{0.11cm}	{\ttfamily \bfseries 11}\\
\vspace{0.11cm}	{\ttfamily \bfseries 12}\\
\vspace{0.11cm}	{\ttfamily \bfseries 13}\\
\vspace{0.11cm}	{\ttfamily \bfseries 14}\\
\vspace{0.11cm}	{\ttfamily \bfseries 15}\\
\vspace{0.11cm}	{\ttfamily \bfseries 16}\\
\vspace{0.11cm}	{\ttfamily \bfseries 17}\\
\vspace{0.33cm} {\ttfamily \bfseries 18}\\
\vspace{0.11cm}	{\ttfamily \bfseries 19}\\
\vspace{0.11cm}	{\ttfamily \bfseries 20}
\end{flushleft}\end{minipage}
\end{table}

\subsection{AcceptEventAction}
AcceptEventActions have different behaviors when they have inputs or not, and therefore, the node need more rules to be specified. A node without pins are activated when the activity is invoked. 

\textbf{Token consumption and invocation for a node with inpins:}
The node is invokated when the required amount of tokens has been offered.

\begin{table}[!h] \begin{minipage}{0.98\linewidth}\centering \footnotesize
\begin{equation*} 
\frac{\parbox{4.7in} {
\vspace{-0.4cm}
 $a \in A, n \in node(a),type(n) = AcceptEventAction, $\\
$S_a(a)=\langle executing, P_s, P_n\rangle, S_n(n)= idle, $\\
$ inpin(n) \neq \emptyset, $\\
$ f_{tl} \in F_{tl}(inpin(n)),\forall p \in inpin(n). transfer(f_{tl}(p)) \neq \emptyset,$\\
 $ \lbrace  p_1, ..., p_i \rbrace = inpin(n),$\\
 $ \lbrace  q_1, ..., q_i \rbrace = \lbrace source(f_{tl}(p_k))| \forall 1 \leq k \leq i\rbrace,$\\
 $ \lbrace  Vc_1, ..., Vc_i\rbrace = \lbrace transfer(f_{tl}(p)) | \forall p \in inpin(n)\rbrace,$\\
 $ \lbrace  V_{q_1}, ..., V_{q_i}\rbrace = \lbrace V | \forall 1 \leq k \leq i. S_{th}(q_k)=V \rbrace$
}}{ \parbox{4.7in} {
$\langle   S_n, S_a,  S_{th}, S_{\Sigma}  \rangle \overset{i(n)} \twoheadrightarrow \langle S_n [ n \mapsto \langle executing, \emptyset \rangle], S_a, $\\
$S_{th}[ q_1 \mapsto V_{q_1}  \rceil   Vc_1]...[ q_i \mapsto V_{q_i}  \rceil   Vc_i],  S_{\Sigma}\rangle$
 }}
\end{equation*}
\end{minipage}
\begin{minipage}{0.01\linewidth}\scriptsize \begin{flushleft}
\vspace{-0.25cm} {\ttfamily \bfseries 1}\\
\vspace{0.11cm}	{\ttfamily \bfseries 2}\\
\vspace{0.11cm}	{\ttfamily \bfseries 3}\\
\vspace{0.11cm}	{\ttfamily \bfseries 4}\\
\vspace{0.11cm}	{\ttfamily \bfseries 5}\\
\vspace{0.11cm}	{\ttfamily \bfseries 6}\\
\vspace{0.11cm}	{\ttfamily \bfseries 7}\\
\vspace{0.11cm} {\ttfamily \bfseries 8}\\
\vspace{0.33cm}	{\ttfamily \bfseries 9}\\
\vspace{0.11cm}	{\ttfamily \bfseries 10}
\end{flushleft}\end{minipage}
\end{table}

\clearpage
\newpage

\textbf{Receiving events, termination and offering tokens:}
This rule is only applicable for node that have inpin. Once the node has received the correspoding event, the execution of the node is terminated by this rule. Thus, the node is not able to receive more events. 

\begin{table}[!h] \begin{minipage}{0.98\linewidth}\centering \footnotesize
\begin{equation*} 
\frac{\parbox{4.7in} {
\vspace{-0.4cm}
 $a \in A, n \in node(a),type(n) = AcceptEventAction, S_a(a)=\langle executing, P_s, P_n\rangle,  $\\
$ S_n(n)= \langle executing, \emptyset \rangle, inpin(n) \neq \emptyset, $\\
$  \langle \sigma, v_e \rangle \in V. event(n)=\sigma, $\\
$ r = result(n), \lbrace  p_1, ..., p_i \rbrace = outpin(n)- r,$\\
$ \lbrace  V_{p_1}, ..., V_{p_i}\rbrace = \lbrace V | \forall 1 \leq k \leq i. S_{th}(p_k)=V \rbrace,$\\
$  V_r = \lbrace V |  S_{th}(r)=V \rbrace,$\\
$|V_r| + 1 \leq  upperbound(r) \wedge \forall 1 \leq k \leq i. |V_{p_k}| + 1 \leq upperbound(p_k)$
}}{ \parbox{4.7in} {
$\langle  S_n, S_a,  S_{th}, S_{\Sigma}  \rangle \overset{t(n)} \twoheadrightarrow \langle  S_n [ n \mapsto idle], S_a, S_{th}[r \mapsto V_r \cup \lbrace v_e \rbrace ]$\\
$[ p_1 \mapsto V_{p_1} \cup \lbrace ControlToken \rbrace]...[ p_i \mapsto V_{p_i} \cup \lbrace ControlToken \rbrace], S_{\Sigma} \rangle$
 }}
\end{equation*}
\end{minipage}
\begin{minipage}{0.01\linewidth}\scriptsize \begin{flushleft}
\vspace{-0.25cm} {\ttfamily \bfseries 1}\\
\vspace{0.11cm}	{\ttfamily \bfseries 2}\\
\vspace{0.11cm}	{\ttfamily \bfseries 3}\\
\vspace{0.11cm}	{\ttfamily \bfseries 4}\\
\vspace{0.11cm}	{\ttfamily \bfseries 5}\\
\vspace{0.11cm}	{\ttfamily \bfseries 6}\\
\vspace{0.11cm}	{\ttfamily \bfseries 7}\\
\vspace{0.33cm} {\ttfamily \bfseries 8}\\
\vspace{0.11cm}	{\ttfamily \bfseries 9}
\end{flushleft}\end{minipage}
\end{table}

\textbf{Receiving events and offering tokens without termination:}
In contrast to previous rule, the execution of AcceptEventActions is not terminated after receiving a event. These node are executing as long as the activity is executing.

\begin{table}[!h] \begin{minipage}{0.98\linewidth}\centering \footnotesize
\begin{equation*} 
\frac{\parbox{4.7in} {
\vspace{-0.4cm}
 $a \in A, n \in node(a),type(n) = AcceptEventAction, S_a(a)=\langle executing, P_s, P_n\rangle,  $\\
$ S_n(n)= \langle executing, \emptyset \rangle, inpin(n) = \emptyset, $\\
$  \langle \sigma, v_e \rangle \in V. event(n)=\sigma, $\\
$ r = result(n), \lbrace  p_1, ..., p_i \rbrace = outpin(n)- r,$\\
$ \lbrace  V_{p_1}, ..., V_{p_i}\rbrace = \lbrace V | \forall 1 \leq k \leq i. S_{th}(p_k)=V \rbrace,$\\
$  V_r = \lbrace V |  S_{th}(r)=V \rbrace,$\\
$|V_r| + 1 \leq  upperbound(r) \wedge \forall 1 \leq k \leq i. |V_{p_k}| + 1 \leq upperbound(p_k)$
}}{ \parbox{4.7in} {
$\langle  S_n, S_a,  S_{th}, S_{\Sigma}  \rangle \overset{t(n)} \twoheadrightarrow \langle  S_n, S_a, S_{th}[r \mapsto V_r \cup \lbrace v_e \rbrace ]$\\
$[ p_1 \mapsto V_{p_1} \cup \lbrace ControlToken \rbrace]...[ p_i \mapsto V_{p_i} \cup \lbrace ControlToken \rbrace], S_{\Sigma} \rangle $
 }}
\end{equation*}
\end{minipage}
\begin{minipage}{0.01\linewidth}\scriptsize \begin{flushleft}
\vspace{-0.25cm} {\ttfamily \bfseries 1}\\
\vspace{0.11cm}	{\ttfamily \bfseries 2}\\
\vspace{0.11cm}	{\ttfamily \bfseries 3}\\
\vspace{0.11cm}	{\ttfamily \bfseries 4}\\
\vspace{0.11cm}	{\ttfamily \bfseries 5}\\
\vspace{0.11cm}	{\ttfamily \bfseries 6}\\
\vspace{0.11cm}	{\ttfamily \bfseries 7}\\
\vspace{0.33cm} {\ttfamily \bfseries 8}\\
\vspace{0.11cm}	{\ttfamily \bfseries 9}
\end{flushleft}\end{minipage}
\end{table}

\clearpage
\newpage

\subsection{SendSignalAction}

\textbf{Token consumption, invocation, sending of the signal:}
Once the invocation requirements of the node are satisfied, a event is saved in the event pool.

\begin{table}[!h] \begin{minipage}{0.98\linewidth}\centering \footnotesize
\begin{equation*} 
\frac{\parbox{4.7in} {
\vspace{-0.4cm}
$a \in A, n \in node(a),type(n) = SendSignalAction,$\\
$S_a(a)=\langle executing, P_s, P_n\rangle, S_n(n)= idle, $\\
$ f_{tl} \in F_{tl}(inpin(n)),\forall p \in inpin(n). transfer(f_{tl}(p)) \neq \emptyset,$\\
 $ \lbrace  Vc_1, ..., Vc_i\rbrace = \lbrace transfer(f_{tl}(p)) | \forall p \in inpin(n)\rbrace,$\\
 $ \lbrace  q_1, ..., q_i \rbrace = \lbrace source(f_{tl}(p_k))| \forall 1 \leq k \leq i\rbrace,$\\
 $ \lbrace  V_{q_1}, ..., V_{q_i}\rbrace = \lbrace V | \forall 1 \leq k \leq i. S_{th}(q_k)=V \rbrace$
}}{ \parbox{4.7in} {
$\langle   S_n, S_a,  S_{th}, S_{\Sigma}  \rangle \overset{i(n)} \twoheadrightarrow \langle S_n [ n \mapsto \langle executing, \emptyset \rangle], S_a, S_{th}[ q_1 \mapsto V_{q_1} \rceil  Vc_1]...$\\
$[ q_i \mapsto V_{q_i} \rceil  Vc_i],  S_{\Sigma} \cup \langle event(n), sendData( Vc_1, ..., Vc_i)\rangle\rangle$
 }}
\end{equation*}
\end{minipage}
\begin{minipage}{0.01\linewidth}\scriptsize \begin{flushleft}
\vspace{-0.25cm} {\ttfamily \bfseries 1}\\
\vspace{0.11cm}	{\ttfamily \bfseries 2}\\
\vspace{0.11cm}	{\ttfamily \bfseries 3}\\
\vspace{0.11cm}	{\ttfamily \bfseries 4}\\
\vspace{0.11cm}	{\ttfamily \bfseries 5}\\
\vspace{0.11cm}	{\ttfamily \bfseries 6}\\
\vspace{0.33cm}	{\ttfamily \bfseries 7}\\
\vspace{0.11cm} {\ttfamily \bfseries 8}
\end{flushleft}\end{minipage}
\end{table}

\textbf{Execution Termination and token offering:}
\begin{table}[!h] \begin{minipage}{0.98\linewidth}\centering \footnotesize
\begin{equation*} 
\frac{\parbox{4.7in} {
\vspace{-0.4cm}
 $a \in A, n \in node(a), type(n) = SendSignalAction,$\\
 $ S_a(a)=\langle executing, P_s, P_n \rangle, S_n(n)= \langle executing, \emptyset \rangle,$\\
$\lbrace  q_1, ..., q_i \rbrace = outpin(n),$\\
$ \lbrace V_{q_1}, ..., V_{q_i} \rbrace = \lbrace  V | \forall q \in outpin(n).S_{th}(q)=V\rbrace, $\\
$ \forall 1 \leq k. |V{q_k}| < upperbound (q_k) $
}}{ \parbox{4.7in} {
$\langle  S_n, S_a,  S_{th}, S_{\Sigma}  \rangle \overset{t(n)} \twoheadrightarrow \langle  S_n [ n \mapsto idle], S_a, S_{th}[ q_1 \mapsto V_{q_1} \cup {ControlToken}]...$\\
$[ q_i \mapsto =V_{q_i} \cup {ControlToken}], S_{\Sigma} \rangle$
 }}
\end{equation*}
\end{minipage}
\begin{minipage}{0.01\linewidth}\scriptsize \begin{flushleft}
\vspace{-0.25cm} {\ttfamily \bfseries 1}\\
\vspace{0.11cm}	{\ttfamily \bfseries 2}\\
\vspace{0.11cm}	{\ttfamily \bfseries 3}\\
\vspace{0.11cm}	{\ttfamily \bfseries 4}\\
\vspace{0.11cm}	{\ttfamily \bfseries 5}\\
\vspace{0.33cm}	{\ttfamily \bfseries 6}\\
\vspace{0.11cm}	{\ttfamily \bfseries 7}
\end{flushleft}\end{minipage}
\end{table}

\subsection{CallBehaviorAction}

Multiple hierarchical levels can be built by using \textit{CallBehaviorActions}. This type of node invokes an activity by making tokens available to the activity that are offered to the node. If the node is $synchronous$, the execution of the node terminates only when the corresponding activity execution is finished (see Appendix). The results of the execution of the activity are offered in the outputs of the  node. Furthermore, $pins$ of the nodes inherit properties of the $APN$ of the activity, such as, parameter set, $streaming$, and $exception$. 

\textbf{Token consumption and invocation with non-streaming and streaming parameters:}

While tokens can be consumed by streaming parameters during the execution of the activity, tokens offered to non-streaming parameter are required to invoke the activity and are only consuming at the invocation point \cite[p. 410]{UML241}. $P_{req}$ defines the set of pins of the node that are required for invoking the activity (Line 4). $m_p$ maps $pins$ of a $CallBehaviorAction$ to the $APN$ of the  activity. Note that all required pins belong to the same $PS$ of the activity. If multiple $PS$ have enough tokens to start the invocation, one of them is chosen non-deterministically. This rule is applicable only if all required pins have been offered enough tokens (Line 5). $P_{cons}$ contains streaming and non-streaming pins that have been offered tokens belonging to the chosen $PS$ (Line 6). Tokens offered to these pins are consumed by the invocation  (Lines 8,14). The sequence of tokens to consume is defined by $Vc_i$ (Line 10). The $APN$ are updated by adding the consumed tokens (Line 14). This rule invokes the node and transfer tokens to the activity. An extra rule is used to invoke the activity.

The execution of an activity starts when any input has a token. Requirements of the activity are evaluated in the invocation of the \textit{CallBehaviorAction}. Tokens flow and events reception is started by invocating \textit{IntialNodes} and \textit{AcceptEventActions}, respectively.

\begin{table}[!h] \begin{minipage}{0.98\linewidth}\centering \footnotesize
\begin{equation*} 
\frac{\parbox{4.7in} {
\vspace{-0.4cm}
  $ a,b \in A,n \in node(a),S_a(a)=\langle executing, P_s, P_n \rangle, S_n(n)= idle,S_a(b)=idle,$\\
 $type(n) = CallBehaviorAction, isSynchronous(n)=true, b=behavior(n), $\\
$PS_b \in PS(b). \exists th \in PS_b, streaming (th) = FALSE,$ \\
$P_{req} = \lbrace p \in inpin(n) | th=m_p(p) \wedge th \in PS_b \wedge streaming(th) = FALSE \rbrace,$\\
$ f_{tl} \in F_{tl}(inpin(n)),\forall p \in P_{req}. transfer(f_{tl}(p)) \neq \emptyset,$\\
$ P_{cons}= \lbrace p \in inpin(n)|m_p(p)  \in PS_b \wedge  transfer(f_{tl}(p)) \neq \emptyset    \rbrace,$\\
 $\lbrace  p_1, ..., p_i \rbrace = P_{cons},$\\
$ \lbrace  q_1, ..., q_i \rbrace = \lbrace source(f_{tl}(p_k))| \forall 1 \leq k \leq i\rbrace,$\\
 $\lbrace  r_1, ..., r_i \rbrace = \lbrace m_p(p_k) | \forall 1 \leq k \leq i \rbrace,$\\
   $ \lbrace  Vc_1, ..., Vc_i\rbrace = \lbrace transfer(f_{tl}(p)) | \forall p \in P_{cons}\rbrace,$\\
  $\forall 1 \leq k \leq i.V_{q_k}=S_{th}(q_k),$\\
$\forall 1 \leq k \leq i.V_{r_k}=S_{th}(r_k)$ 
}}{ \parbox{4.7in} {
$\langle   S_n, S_a,  S_{th}, S_{\Sigma}  \rangle \overset{i(n)} \twoheadrightarrow \langle S_n [ n \mapsto \langle executing, \emptyset \rangle], S_a, $\\
$S_{th}[ q_1 \mapsto V_{q_1} \rceil Vc_1]...[ q_i \mapsto V_{q_i} \rceil Vc_i] [ r_1 \mapsto V_{r_1} \cup Vc_1]...[ r_i \mapsto V_{r_i} \cup Vc_i], S_{\Sigma}\rangle$
 }}
\end{equation*}
\end{minipage}
\begin{minipage}{0.01\linewidth}\scriptsize \begin{flushleft}
\vspace{-0.25cm} {\ttfamily \bfseries 1}\\
\vspace{0.11cm}	{\ttfamily \bfseries 2}\\
\vspace{0.11cm}	{\ttfamily \bfseries 3}\\
\vspace{0.11cm}	{\ttfamily \bfseries 4}\\
\vspace{0.11cm}	{\ttfamily \bfseries 5}\\
\vspace{0.11cm}	{\ttfamily \bfseries 6}\\
\vspace{0.11cm}	{\ttfamily \bfseries 7}\\
\vspace{0.11cm} {\ttfamily \bfseries 8}\\
\vspace{0.11cm}	{\ttfamily \bfseries 9}\\
\vspace{0.11cm}	{\ttfamily \bfseries 10}\\
\vspace{0.11cm}	{\ttfamily \bfseries 11}\\
\vspace{0.11cm}	{\ttfamily \bfseries 12}\\
\vspace{0.33cm}	{\ttfamily \bfseries 13}\\
\vspace{0.11cm}	{\ttfamily \bfseries 14}
\end{flushleft}\end{minipage}
\end{table}

\textbf{Token consumption and invocation with only streaming parameter:}
If there is only streaming parameter, the \textit{Activity} is invokated if at least one input has been offered a token.

\begin{table}[!h] \begin{minipage}{0.98\linewidth}\centering \footnotesize
\begin{equation*} 
\frac{\parbox{4.7in} {
\vspace{-0.4cm}
 $ a,b \in A,n \in node(a),type(n) = CallBehaviorAction,isSynchronous(n)=true,$\\
 $S_a(a)=\langle executing, P_s, P_n \rangle, S_n(n)= idle, b=behavior(n), S_a(b)=idle, $\\
 $PS_b \in PS(b). \forall th \in PS_b, streaming (th) = TRUE,$\\
 $ f_{tl} \in F_{tl}(inpin(n)). \exists p \in inpin(n). m_p(p) \in PS_b \wedge transfer(f_{tl}(p)) \neq \emptyset,$\\
$ P_{cons}= \lbrace p \in inpin(n)|m_p(p) \in PS_b \wedge transfer(f_{tl}(p)) \neq \emptyset\rbrace,$\\
 $\lbrace  p_1, ..., p_i \rbrace = P_{cons},$\\
 $\lbrace  q_1, ..., q_i \rbrace = \lbrace th \in PS_b | m_p^{-1}(th) \in P_{cons} \rbrace,$\\
 $\lbrace V_{q_1}, ..., V_{q_i} \rbrace  = \lbrace  V_k | \forall 1 \leq k \leq i.S_{th}(q_k)=V_k \rbrace, $\\
  $ \lbrace  V_{r_1}, ..., V_{r_i}\rbrace = \lbrace V | \forall p \in P_{cons}. S_{th}(f_{tl}(p))=V \rbrace,$\\
  $ \lbrace  Vc_1, ..., Vc_i\rbrace = \lbrace transfer(f_{tl}(p)) | \forall p \in P_{cons}\rbrace,$
}}{ \parbox{4.7in} {
$\langle   S_n, S_a,  S_{th}, S_{\Sigma}  \rangle \overset{i(n)} \twoheadrightarrow \langle S_n [ n \mapsto \langle executing, \emptyset \rangle], S_a, S_{th}[ f_{tl} (p_1) \mapsto V_{r_1} \rceil Vc_1]...$\\
$[ f_{tl}(p_i) \mapsto V_{r_i} \rceil Vc_i], [ q_1 \mapsto V_{q_1} \cup Vc_1]...[ q_i \mapsto V_{q_i} \cup Vc_i], S_{\Sigma}\rangle$
 }}
\end{equation*}
\end{minipage}
\begin{minipage}{0.01\linewidth}\scriptsize \begin{flushleft}
\vspace{-0.25cm} {\ttfamily \bfseries 1}\\
\vspace{0.11cm}	{\ttfamily \bfseries 2}\\
\vspace{0.11cm}	{\ttfamily \bfseries 3}\\
\vspace{0.11cm}	{\ttfamily \bfseries 4}\\
\vspace{0.11cm}	{\ttfamily \bfseries 5}\\
\vspace{0.11cm}	{\ttfamily \bfseries 6}\\
\vspace{0.11cm}	{\ttfamily \bfseries 7}\\
\vspace{0.11cm} {\ttfamily \bfseries 8}\\
\vspace{0.11cm}	{\ttfamily \bfseries 9}\\
\vspace{0.11cm}	{\ttfamily \bfseries 10}\\
\vspace{0.33cm}	{\ttfamily \bfseries 11}\\
\vspace{0.11cm}	{\ttfamily \bfseries 12}
\end{flushleft}\end{minipage}
\end{table}

\clearpage
\newpage
\textbf{Token consumption during the execution from streaming parameters:}
Tokens are passed from the \textit{CallBehaviorAction} to the \textit{Activity} during the execution if they have been offered in a input streaming pin.

\begin{table}[!h] \begin{minipage}{0.98\linewidth}\centering \footnotesize
\begin{equation*} 
\frac{\parbox{4.7in} {
\vspace{-0.4cm}
 $ a,b \in A,n \in node(a),type(n) = CallBehaviorAction,b=behavior(n),$\\
 $S_a(a)=\langle executing, P_s', P_n' \rangle, S_n(n)= \langle executing, \emptyset\rangle, S_a(b)=\langle executing, P_s, P_n \rangle,$\\
 $isSynchronous(n)=true,$\\
 $ P_{stream} = \lbrace p \in inpin(n)| th=m_p(p) \wedge th \in P_s \wedge streaming(th) = true \rbrace,$\\
 $ f_{tl} \in F_{tl}(inpin(p)). \exists p \in P_{stream}. transfer(f_{tl}(p)) \neq \emptyset,$\\
$ P_{cons}= \lbrace p \in P_{stream}| transfer(f_{tl}(p)) \neq \emptyset\rbrace,$\\
 $\lbrace  p_1, ..., p_i \rbrace = P_{cons},$\\
 $\lbrace  q_1, ..., q_i \rbrace = \lbrace th \in P_s | m_p^{-1}(th) \in P_{cons} \rbrace,$\\
 $\lbrace  r_1, ..., r_i \rbrace = \lbrace source(f_{tl}(p_k))| \forall 1 \leq k \leq i\rbrace,$\\
 $\forall 1 \leq k \leq i.V_{q_k}=S_{th}(q_k),$\\ 
 $\forall 1 \leq k \leq i.V_{r_k}=S_{th}(r_k),$\\ 
 $\forall 1 \leq k \leq i.Vc_{k}=S_{th}(transfer(f_{tl}(p_k)))$
}}{ \parbox{4.7in} {
$\langle   S_n, S_a,  S_{th}, S_{\Sigma}  \rangle \overset{i(n)} \twoheadrightarrow \langle S_n, S_a[b  \mapsto \langle executing, P_s, P_n - P_{cons} \rangle], $\\
$S_{th}[ r_1\mapsto V_{r_1} \rceil Vc_1]...[ r_i \mapsto V_{r_i} \rceil Vc_i][ q_1 \mapsto V_{q_1} \cup Vc_1]...[ q_i \mapsto V_{q_i} \cup Vc_i], S_{\Sigma}\rangle$
 }}
\end{equation*}
\end{minipage}
\begin{minipage}{0.01\linewidth}\scriptsize \begin{flushleft}
\vspace{-0.25cm} {\ttfamily \bfseries 1}\\
\vspace{0.11cm}	{\ttfamily \bfseries 2}\\
\vspace{0.11cm}	{\ttfamily \bfseries 3}\\
\vspace{0.11cm}	{\ttfamily \bfseries 4}\\
\vspace{0.11cm}	{\ttfamily \bfseries 5}\\
\vspace{0.11cm}	{\ttfamily \bfseries 6}\\
\vspace{0.11cm}	{\ttfamily \bfseries 7}\\
\vspace{0.11cm} {\ttfamily \bfseries 8}\\
\vspace{0.11cm}	{\ttfamily \bfseries 9}\\
\vspace{0.11cm}	{\ttfamily \bfseries 10}\\
\vspace{0.11cm}	{\ttfamily \bfseries 11}\\
\vspace{0.11cm}	{\ttfamily \bfseries 12}\\
\vspace{0.33cm}	{\ttfamily \bfseries 13}\\
\vspace{0.11cm}	{\ttfamily \bfseries 14}
\end{flushleft}\end{minipage}
\end{table}

\textbf{Token offering of streaming outputs:}
Tokens in output streaming paramenters are tranfered to the \textit{CallBehaviorAction} during the execution.

\begin{table}[!h] \begin{minipage}{0.98\linewidth}\centering \footnotesize
\begin{equation*} 
\frac{\parbox{4.7in} {
\vspace{-0.4cm}
$ a,b \in A,n \in node(a),type(n) = CallBehaviorAction,b=behavior(n),$\\
 $S_a(a)=\langle executing, P_s', P_n' \rangle, S_n(n)= \langle executing, \emptyset\rangle, S_a(b)=\langle executing, P_s, P_n \rangle,$\\
 $isSynchronous(n)=true,$\\ 
 $ p \in output(b). S_{th}\neq \emptyset \wedge streaming(th) = true, $ \\
 $ q = m_{p}^{-1}(p),$\\
 $ V_{p} = S_{th}(p), $ \\
 $ V_{q} = S_{th}(q), $ \\
 $|V_{q} \cup V_{p} |\leq upperbound(q) $
}}{ \parbox{4.7in} {
$\langle S_n, S_a,  S_{th}, S_{\Sigma}  \rangle \rightarrow \langle S_n, S_a,S_{th}[p \mapsto \emptyset ][q \mapsto V_q \cup V_p ], S_{\Sigma}\rangle$
 }}
\end{equation*}
\end{minipage}
\begin{minipage}{0.01\linewidth}\scriptsize \begin{flushleft}
\vspace{-0.25cm} {\ttfamily \bfseries 1}\\
\vspace{0.11cm}	{\ttfamily \bfseries 2}\\
\vspace{0.11cm}	{\ttfamily \bfseries 3}\\
\vspace{0.11cm}	{\ttfamily \bfseries 4}\\
\vspace{0.11cm}	{\ttfamily \bfseries 5}\\
\vspace{0.11cm}	{\ttfamily \bfseries 6}\\
\vspace{0.11cm}	{\ttfamily \bfseries 7}\\
\vspace{0.11cm} {\ttfamily \bfseries 8}\\
\vspace{0.33cm}	{\ttfamily \bfseries 9}
\end{flushleft}\end{minipage}
\end{table}

\subsection{Activity}

\textbf{Termination and token offering of an activity:} This rule is applicable when there are no tokens left to process in the activity  or nodes that are been executing (Lines 4,5). Since this rule is for a synchronous call (Line 3), the resulting tokens are offered in the outpins of the corresponding \textit{CallBehaviorAction}, which is also terminated in this rule (Line 15). Note that a \textit{null} token is offered in case that the corresponding outpin is empty by the end of the activity (Lines 9,17). 

\clearpage
\newpage

\begin{table}[!h] \begin{minipage}{0.98\linewidth}\centering \footnotesize
\begin{equation*} 
\frac{\parbox{4.8in} {
\vspace{-0.4cm}
$a,b\in A,n\in node(a),type(n)=CallBehaviorAction,b=behavior(n),$\\ 
 $S_a(a)=\langle executing, P_s', P_n' \rangle, S_n(n)= \langle executing, \emptyset\rangle, S_a(b)=\langle executing, P_s, \emptyset \rangle,$\\
 $isSynchronous(n)=true,$\\
$\forall th \in ( holders(b)-output(b)) .S_{th}(p)=\emptyset,$\\
 $\forall m \in node(b).S_n(m)=idle ,$\\
 $ P_{ocu}= \lbrace th \in output(b) | S_{th}(th) \neq \emptyset  \wedge isExeption(th)=false \rbrace ,$\\
 $P_s \in PS(b).\forall p \in P_{ocu}.p \in P_s$\\  
 $\lbrace  p_1, ..., p_i \rbrace = P= \lbrace p \in P_s | S_{th}\neq \emptyset \rbrace,$\\
 $ \lbrace  r_1, ..., r_j \rbrace = P'=P_s-P,$\\
 $\forall 1 \leq k \leq i.V_{p_k}=S_{th}(p_k),$\\
 $\forall 1 \leq k \leq i.V_{q_k}=S_{th}(m_p(q_k)),$\\
 $\forall 1 \leq k \leq j.V_{r_k}=S_{th}(m_p(r_k)),$\\
$ \forall q_k \in m_p^{-1}(P). | V_{q_k} \cup V_{p_k} |\leq upperbound(q_k) ,$\\
$ \forall r_k \in m_p^{-1}(P'). |V_{r_k}| + 1\leq upperbound(r_k)$
}}{ \parbox{4.7in} {
$\langle  S_n, S_a,  S_{th}, S_{\Sigma}  \rangle \overset{t(n)} \twoheadrightarrow \langle  S_n [ n \mapsto idle], S_a[ b \mapsto idle], S_{th}[p_1 \mapsto \emptyset]...[p_i \mapsto \emptyset]$\\
$[ m_p^{-1}(q_1) \mapsto V_{q_1} \cup V_{p_1}]...m_p^{-1}(p_i) \mapsto V_{p_i} \cup V_{p_i} ]$\\
$[ m_p^{-1}(r_1) \mapsto V_{r_1} \cup \lbrace null \rbrace]...[ m_p^{-1}(r_j) \mapsto V_{r_j} \cup \lbrace null \rbrace ],
 S_{\Sigma} \rangle$
}}
\end{equation*}
\end{minipage}
\begin{minipage}{0.01\linewidth}\scriptsize \begin{flushleft}
\vspace{-0.25cm} {\ttfamily \bfseries 1}\\
\vspace{0.11cm}	{\ttfamily \bfseries 2}\\
\vspace{0.11cm}	{\ttfamily \bfseries 3}\\
\vspace{0.11cm}	{\ttfamily \bfseries 4}\\
\vspace{0.11cm}	{\ttfamily \bfseries 5}\\
\vspace{0.11cm}	{\ttfamily \bfseries 6}\\
\vspace{0.11cm}	{\ttfamily \bfseries 7}\\
\vspace{0.11cm} {\ttfamily \bfseries 8}\\
\vspace{0.11cm}	{\ttfamily \bfseries 9}\\
\vspace{0.11cm}	{\ttfamily \bfseries 10}\\
\vspace{0.11cm}	{\ttfamily \bfseries 11}\\
\vspace{0.11cm}	{\ttfamily \bfseries 12}\\
\vspace{0.11cm}	{\ttfamily \bfseries 13}\\
\vspace{0.11cm}	{\ttfamily \bfseries 14}\\
\vspace{0.33cm}	{\ttfamily \bfseries 15}\\
\vspace{0.11cm}	{\ttfamily \bfseries 16}\\
\vspace{0.11cm}	{\ttfamily \bfseries 17}
\end{flushleft}\end{minipage}
\end{table}

\textbf{Token consumption by streaming parameters during activity execution:} This rule passes tokens to the corresponding activity that are offered to the \textit{CallBehaviorAction}. Only tokens offered in streaming parameters and belonging to the $P_s$ are consumed in this rule (Lines 4,6). APN with incoming tokens that have not been set before are removed from the set $P_n$, since a empty $P_n$ is required for the termination of the activity (Line 13).

\begin{table}[!h] \begin{minipage}{0.98\linewidth}\centering \footnotesize
\begin{equation*} 
\frac{\parbox{4.7in} {
\vspace{-0.4cm}
$ a \in A,S_a(a)=idle, $\\
 $ P_{ocu}= \lbrace th \in APN(a) | S_{th}(th) \neq \emptyset \rbrace, $\\
 $ P_s \in PS(a).\forall p \in P_{ocu}.p \in P_s \wedge P_{ocu} \neq \emptyset,$\\
  $ \lbrace  n_1, ..., n_i \rbrace = \lbrace n \in node(a) | type(n) = InitialNode  \rbrace,$\\
  $ \lbrace  m_1, ..., m_j \rbrace = \lbrace m \in node(a) | type(m) = AcceptEventAction \wedge inpin(m) = \emptyset \rbrace$
}}{ \parbox{4.7in} {
$\langle  S_n, S_a,  S_{th}, S_{\Sigma}\rangle \overset{i(n)} \twoheadrightarrow \langle S_n [m_1 \mapsto \langle executing, \emptyset \rangle]...[m_j \mapsto \langle executing, \emptyset \rangle],$\\
$ S_a [a \mapsto \langle executing, P_s, P_s - P_{ocu} \rangle],$\\
$S_{th}[ outpin(n_1) \mapsto \lbrace CT \rbrace]...[ outpin(n_j) \mapsto \lbrace CT \rbrace], S_{\Sigma}\rangle$\\
 }}
\end{equation*}
\end{minipage}
\begin{minipage}{0.01\linewidth}\scriptsize \begin{flushleft}
\vspace{-0.25cm} {\ttfamily \bfseries 1}\\
\vspace{0.11cm}	{\ttfamily \bfseries 2}\\
\vspace{0.11cm}	{\ttfamily \bfseries 3}\\
\vspace{0.11cm}	{\ttfamily \bfseries 4}\\
\vspace{0.11cm}	{\ttfamily \bfseries 5}\\
\vspace{0.33cm}	{\ttfamily \bfseries 6}\\
\vspace{0.11cm}	{\ttfamily \bfseries 7}\\
\vspace{0.11cm} {\ttfamily \bfseries 8}
\end{flushleft}\end{minipage}
\end{table}

\vspace{-0.5cm}
\textbf{OutParameter:}
This rule transfers tokens that have been offered to \textit{APN}, which are necessary to the termination of the \textit{Activity}.
\vspace{-0.5cm}

\begin{table}[!h] \begin{minipage}{0.98\linewidth}\centering \footnotesize
\begin{equation*} 
\frac{\parbox{4.7in} {
\vspace{-0.4cm}
 $ a \in A. S_a(a)=\langle executing, P_s, P_n \rangle,$\\
 $  e \in edge(a). e = \langle s,t,g\rangle \wedge t \in APN(a) \wedge s \in pin(a) \wedge isExeption(t)=FALSE \wedge transfer(e) \neq \emptyset,$\\
 $Vc=transfer(e),\\$
 $Vs = \lbrace V | S_{th}(source(e))=V \rbrace,$\\
 $Vt = \lbrace V | S_{th}(target(e))=V \rbrace$
}}{ \parbox{4.7in} {
$\langle S_n, S_a,  S_{th}, S_{\Sigma}  \rangle \rightarrow \langle S_n, S_a,S_{th}[source(e) \mapsto Vs \rceil Vc ]$\\
$[target(e) \mapsto Vt \cup Vc ], S_{\Sigma}\rangle$
 }}
\end{equation*}
\end{minipage}
\begin{minipage}{0.01\linewidth}\scriptsize \begin{flushleft}
\vspace{-0.25cm} {\ttfamily \bfseries 1}\\
\vspace{0.11cm}	{\ttfamily \bfseries 2}\\
\vspace{0.11cm}	{\ttfamily \bfseries 3}\\
\vspace{0.11cm}	{\ttfamily \bfseries 4}\\
\vspace{0.11cm}	{\ttfamily \bfseries 5}\\
\vspace{0.11cm}	{\ttfamily \bfseries 6}\\
\vspace{0.33cm}	{\ttfamily \bfseries 7}\\
\vspace{0.11cm} {\ttfamily \bfseries 8}
\end{flushleft}\end{minipage}
\end{table}

\clearpage
\newpage

\subsection{Exceptions}
\textbf{Throwing an exception:}
An \textit{Activity} is in an exception if an exception parameter has received a token. 
\vspace{-0.2cm}
\begin{table}[!h] \begin{minipage}{0.98\linewidth}\centering \footnotesize
\begin{equation*} 
\frac{\parbox{4.7in} {
\vspace{-0.4cm}
 $ a \in A. S_a(a)=\langle executing, P_s, P_n \rangle,$\\
 $  e \in edge(a). e = \langle s,t,g\rangle \wedge t \in APN(a) \wedge s \in pin(a) \wedge isExeption(t)=TRUE \wedge transfer(e) \neq \emptyset,$\\
 $V_e=transfer(e),$\\
 $ \lbrace  p_1, ..., p_i \rbrace = pin(a),$\\
 $ \lbrace  n_1, ..., n_j \rbrace = node(a)$
}}{ \parbox{4.7in} {
$\langle S_n, S_a,  S_{th}, S_{\Sigma}  \rangle \rightarrow \langle S_n[n_1 \mapsto idle]...[n_j \mapsto idle],S_a[a \mapsto exception(V_e)],$\\
$ S_{th}[  p_1 \mapsto \emptyset]...[ p_i \mapsto \emptyset],S_{\Sigma}\rangle$
 }}
\end{equation*}
\end{minipage}
\begin{minipage}{0.01\linewidth}\scriptsize \begin{flushleft}
\vspace{-0.25cm} {\ttfamily \bfseries 1}\\
\vspace{0.11cm}	{\ttfamily \bfseries 2}\\
\vspace{0.11cm}	{\ttfamily \bfseries 3}\\
\vspace{0.11cm}	{\ttfamily \bfseries 4}\\
\vspace{0.11cm}	{\ttfamily \bfseries 5}\\
\vspace{0.11cm}	{\ttfamily \bfseries 6}\\
\vspace{0.33cm}	{\ttfamily \bfseries 7}\\
\vspace{0.11cm} {\ttfamily \bfseries 8}
\end{flushleft}\end{minipage}
\end{table}

\textbf{Inkovation of the handler of an exception:}
After the exception in throwed, the exception handler is invoked if there is any.
\vspace{-0.2cm}
\begin{table}[!h] \begin{minipage}{0.98\linewidth}\centering \footnotesize
\begin{equation*} 
\frac{\parbox{4.7in} {
\vspace{-0.4cm}
$a,b \in A, n \in node(a),  S_n(n)= \langle idle,\emptyset \rangle, S_a(a)=\langle executing, P_s, P_n\rangle, $\\
$S_a(b)=\langle exception, v_e \rangle,isSynchronous(n)=true, S_n(n)= \langle idle,\emptyset \rangle,$\\
$n \in handler(b).typeException(n)= \delta(v_e),$\\
$ \forall q \in outpin(h).S_{th}= \emptyset$\\
 $p=inpin(h)$
}}{ \parbox{4.7in} {
$\langle S_n, S_a,  S_{th}, S_{\Sigma}  \rangle \rightarrow \langle S_n,S_a, S_{th}[  p \mapsto v_e],S_{\Sigma}\rangle$
 }}
\end{equation*}
\end{minipage}
\begin{minipage}{0.01\linewidth}\scriptsize \begin{flushleft}
\vspace{-0.25cm} {\ttfamily \bfseries 1}\\
\vspace{0.11cm}	{\ttfamily \bfseries 2}\\
\vspace{0.11cm}	{\ttfamily \bfseries 3}\\
\vspace{0.11cm}	{\ttfamily \bfseries 4}\\
\vspace{0.11cm}	{\ttfamily \bfseries 5}\\
\vspace{0.33cm}	{\ttfamily \bfseries 6}
\end{flushleft}\end{minipage}
\end{table}

\textbf{Propagation of the exception in case of no handler:}
In case that there is no exception handler, the exception is propagate to the next hierarchical level. 
\vspace{-0.4cm}
\begin{table}[!h] \begin{minipage}{0.98\linewidth}\centering \footnotesize
\begin{equation*} 
\frac{\parbox{4.7in} {
\vspace{-0.4cm}
$a,b \in A, n\in node(a),type(n)=CallBehaviorAction,b=behavior(n), $\\
$S_a(b)=\langle exception, v_e \rangle,S_n(n)= \langle executing,\emptyset \rangle, S_a(a)=\langle executing, P_s, P_n\rangle,$\\
$isSynchronous(n)=true,$\\
$  \not \exists n \in handler(b).typeException(n)= \delta(v_e),$\\
 $ \lbrace  p_1, ..., p_i \rbrace = pin(a),$\\
 $ \lbrace  n_1, ..., n_j \rbrace = node(a)$
}}{ \parbox{4.7in} {
$\langle S_n, S_a,  S_{th}, S_{\Sigma}  \rangle \rightarrow \langle S_n[n_1 \mapsto idle]...[n_j \mapsto idle],S_a[a \mapsto exception(V_e)][b \mapsto idle],$\\
$ S_{th}[  p_1 \mapsto \emptyset]...[ p_i \mapsto \emptyset],S_{\Sigma}\rangle$
 }}
\end{equation*}
\end{minipage}
\begin{minipage}{0.01\linewidth}\scriptsize \begin{flushleft}
\vspace{-0.25cm} {\ttfamily \bfseries 1}\\
\vspace{0.11cm}	{\ttfamily \bfseries 2}\\
\vspace{0.11cm}	{\ttfamily \bfseries 3}\\
\vspace{0.11cm}	{\ttfamily \bfseries 4}\\
\vspace{0.11cm}	{\ttfamily \bfseries 5}\\
\vspace{0.11cm}	{\ttfamily \bfseries 6}\\
\vspace{0.11cm}	{\ttfamily \bfseries 7}\\
\vspace{0.3cm} {\ttfamily \bfseries 8}\\
\vspace{0.11cm}	{\ttfamily \bfseries 9}
\end{flushleft}\end{minipage}
\end{table}

\textbf{Termination of the exception due to the asynchronous call:}
If the \textit{Activity} has been asynchronously invocated and the exception has not been hadler, the exception is deleted.
\vspace{-0.5cm}
\begin{table}[!h] \begin{minipage}{0.98\linewidth}\centering \footnotesize
\begin{equation*} 
\frac{\parbox{4.7in} {
\vspace{-0.4cm}
$a,b \in A, n \in node(a),  S_n(n)= \langle idle,\emptyset \rangle, S_a(a)=\langle executing, P_s, P_n\rangle,$\\
$ S_a(b)=\langle exception, v_e \rangle, isSynchronous(n)=false$
}}{ \parbox{4.7in} {
$\langle S_n, S_a,  S_{th}, S_{\Sigma}  \rangle \rightarrow \langle S_n,S_a[b\mapsto idle], S_{th},S_{\Sigma}\rangle$
 }}
\end{equation*}
\end{minipage}
\begin{minipage}{0.01\linewidth}\scriptsize \begin{flushleft}
\vspace{-0.25cm} {\ttfamily \bfseries 1}\\
\vspace{0.11cm}	{\ttfamily \bfseries 2}\\
\vspace{0.11cm}	{\ttfamily \bfseries 3}\\
\vspace{0.33cm}	{\ttfamily \bfseries 4}
\end{flushleft}\end{minipage}
\end{table}

\clearpage
\newpage

\textbf{Transfering outgoing tokens from exception handler:}
After a handler terminates the execution, its results are tranfered to the \textit{Activity} that throwed the exception.

\begin{table}[!h] \begin{minipage}{0.98\linewidth}\centering \footnotesize
\begin{equation*} 
\frac{\parbox{4.7in} {
\vspace{-0.4cm}
$a,b \in A, n,m \in node(a),  S_n(n)= \langle idle,\emptyset \rangle, S_a(a)=\langle executing, P_s, P_n\rangle, S_a(b)=\langle exception, v_e \rangle,$\\
$isSynchronous(n)=true, type(m)=CallBehaviorAction,b=behavior(m),$\\
$n \in handler(b).typeException(n)= \delta(v_e),$\\
$ \exists p \in outpin(n).S_{th} \neq \emptyset$\\
$ \lbrace  p_1, ..., p_i \rbrace = outpin(a),$\\
$\forall 1 \leq k \leq i.q_k=m_e(p_k),$\\
$\forall 1 \leq k \leq i.V_{p_k}=S_{th}(p_k),$\\
$\forall 1 \leq k \leq i.V_{q_k}=S_{th}(q_k),$\\
$\lbrace  r_1, ..., r_j \rbrace = \lbrace r \in outpin(m) | \not \exists q_k, q_k = r\rbrace,$

}}{ \parbox{4.7in} {
$\langle S_n, S_a,  S_{th}, S_{\Sigma}  \rangle \rightarrow \langle S_n,S_a [ b \mapsto idle], S_{th}[p_1 \mapsto \emptyset]...[p_i \mapsto \emptyset][ q_1 \mapsto V_{q_1} \cup V_{p_1}]...[q_i \mapsto V_{p_i} \cup V_{p_i} ]$\\
$[ r_1 \mapsto \lbrace CT \rbrace]...[r_j \mapsto \lbrace CT \rbrace], S_{\Sigma}\rangle$
 }}
\end{equation*}
\end{minipage}
\begin{minipage}{0.01\linewidth}\scriptsize \begin{flushleft}
\vspace{-0.25cm} {\ttfamily \bfseries 1}\\
\vspace{0.11cm}	{\ttfamily \bfseries 2}\\
\vspace{0.11cm}	{\ttfamily \bfseries 3}\\
\vspace{0.11cm}	{\ttfamily \bfseries 4}\\
\vspace{0.11cm}	{\ttfamily \bfseries 5}\\
\vspace{0.11cm}	{\ttfamily \bfseries 6}\\
\vspace{0.11cm}	{\ttfamily \bfseries 7}\\
\vspace{0.11cm} {\ttfamily \bfseries 8}\\
\vspace{0.33cm}	{\ttfamily \bfseries 9}
\end{flushleft}\end{minipage}
\end{table}

\section{Semantics extensions}

Semantics extensions refer to the extension of the core semantics in order to define a specific interpretation of  UML activities. 
Depending on the extension, possible behaviors given by the core semantics may be reduced or enlarged, which can affect the consistency with the standard. This section shows some examples that extend the semantics described in previous section. Extensions in the execution time and single core processing are based on the profile DMOSES \cite{Daw2009b}, which adds information to UML models regarding execution time, parallelism and priority in order to generate software for embedded systems. Previous work \cite{Daw2015b} presented a approach to determine if an extension to the semantics is consistent with the standard that is based on the simulation ordering of Park.

\subsection{Execution time}
For real-time systems, violations of time deadlines can lead to significant errors in the system. Therefore, information about the duration of the execution of nodes has to be modeled in order to be able to verify timing constraints. Since the UML standard does not specify the duration of the execution of nodes, timing information can be added to the model using profiling. The following inference rule $exeTime(n)$ defines the time that has elapsed since the invocation of the action ($i(n)$), where the clock is started. The semantics of the clocks process implements the time increasing and ensures a correct termination. This rule adds a new label to the semantics that does not exist in the reference semantics. Note that a set of clocks is added to the state of the system and the status $ready$ is added to the node. The conclusion of the rule stops the timer corresponding to the node that enables a termination rule.

\begin{table}[h] \begin{minipage}{0.98\linewidth}\centering \footnotesize
\begin{equation*}\frac{\parbox{4.3in} {
\vspace{-0.4cm}
$a \in A, n \in node(a),type(n) = Action,S_a(a)=\langle executing, P_s, P_n\rangle,$\\
$S_n(n)= \langle executing,f_{in} \rangle, $\\
$  C(n) \geq executionTime(n)$
}}{ \parbox{4.3in} {
$\langle  S_n, S_a,  S_{th}, S_{\Sigma} ,C \rangle \overset{exeTime(n)} \twoheadrightarrow \langle S_n [ n \mapsto \langle ready, f_{in}\rangle], S_a,  S_{th}, S_{\Sigma}, C [n \mapsto \bot ]\rangle$
 }}
\end{equation*}
\end{minipage}
\begin{minipage}{0.01\linewidth}\scriptsize \begin{flushleft}
\vspace{-0.25cm} {\ttfamily \bfseries 1}\\
\vspace{0.11cm}	{\ttfamily \bfseries 2}\\
\vspace{0.11cm}	{\ttfamily \bfseries 3}\\
\vspace{0.3cm} {\ttfamily \bfseries 4}
\end{flushleft}\end{minipage}
\end{table}

\subsection{Non-preemptive single core}

Activities implemented by non-preemptive single core processors are a good example of how a behavior can be limited by the target platform. Although multiple nodes can be being executed at the same time in a UML activity, a single core processor can only execute one at the time. The following premises are added to all invocation rules ($i(n)$) to specify non-preemptive single core processing behavior. These premises limit the applicability of the invocation rules since no node can be invoked if there is another being executed.

\begin{equation*}
\forall a \in A.\not \exists m \in node(a). m \neq n \wedge S_n(m)=\langle executing, f_{in}' \rangle
\end{equation*}

\subsection{Implementations of token consumption}
\label{sec:ImpTokCons}

\begin{figure}[h]
\centering
\subfigure[Activity A] {\includegraphics[width=0.4\textwidth]{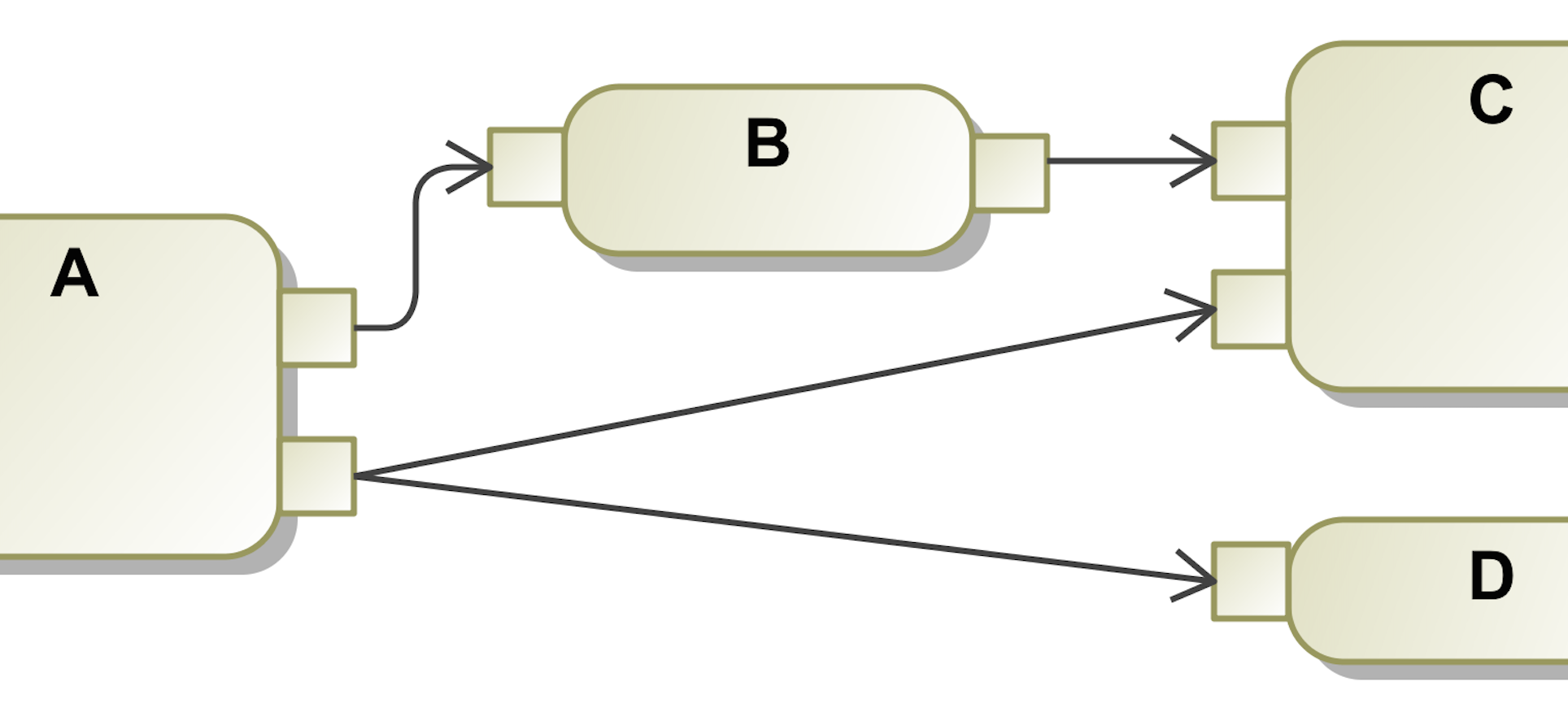} \label{fig:ActA}}
\hspace{2cm}
\subfigure[Reference Semantics] {\includegraphics[width=0.34\textwidth]{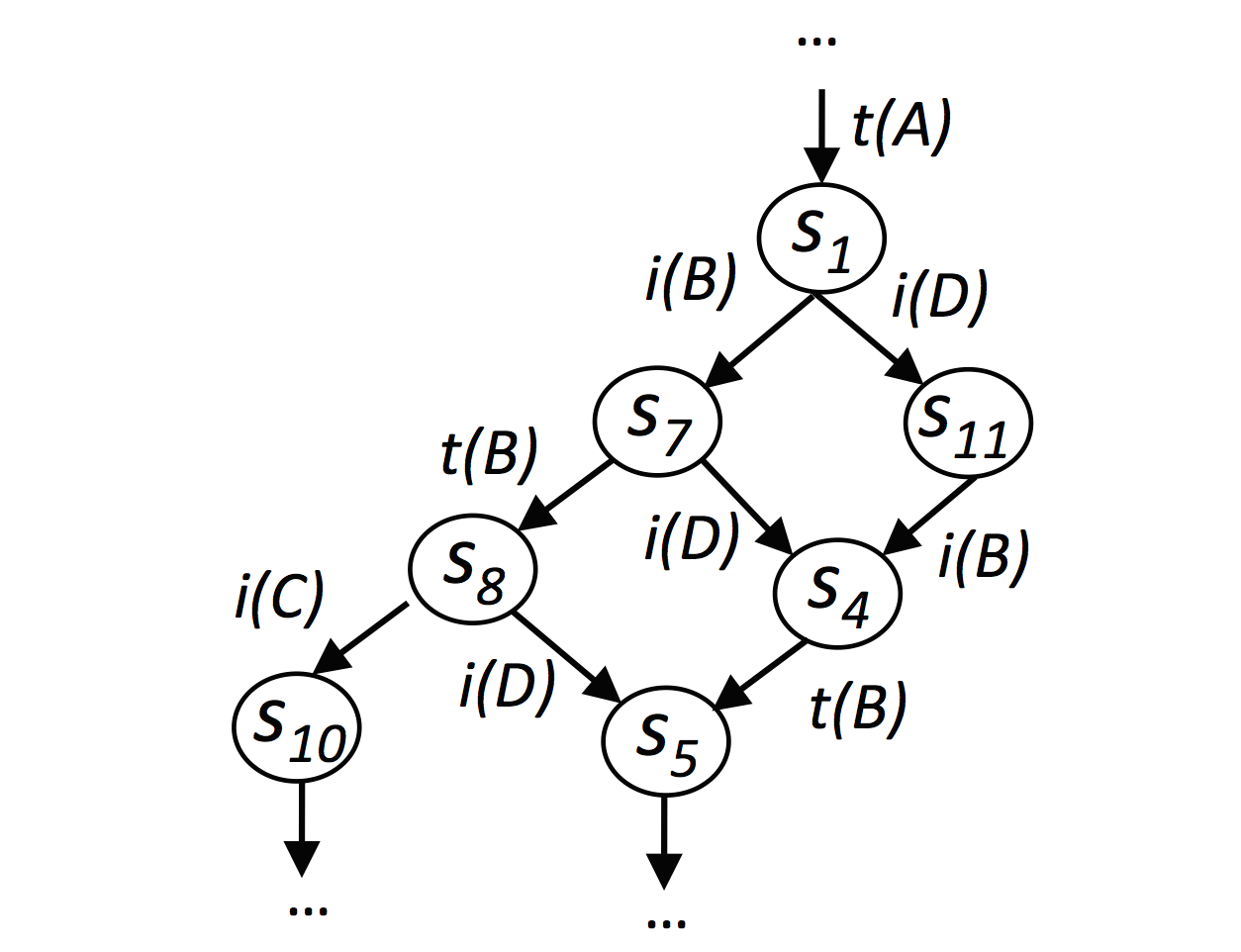} \label{fig:CoreSem}}
\subfigure[Variation 1] {\includegraphics[width=0.18\textwidth]{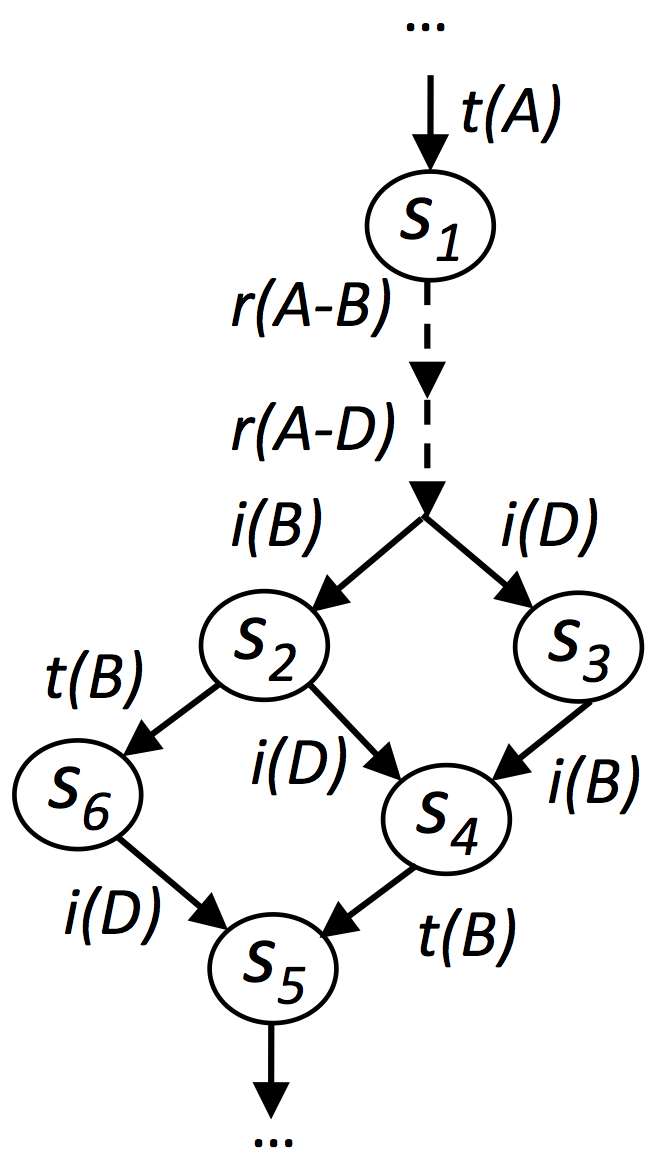} \label{fig:Var1}}
\hspace{2cm}
\subfigure[Variation 2] 
{\includegraphics[width=0.45\textwidth]{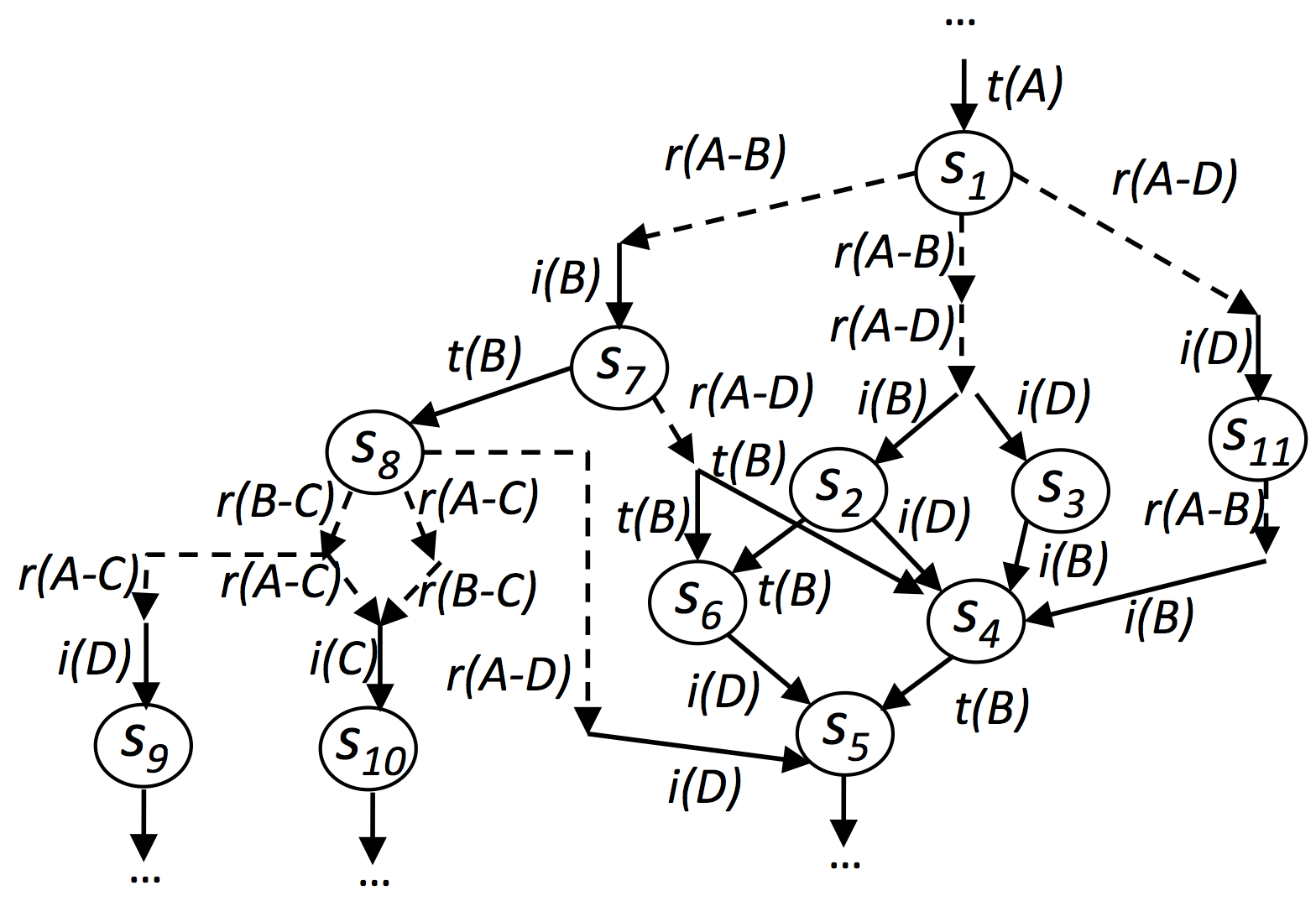} \label{fig:Var2}}
\caption{The implementation of token consumption can reduce or increase the number of possible behaviors}
\label{fig:TokenC}
\end{figure}

The UML standard defines that tokens can be transferred only if they can be immediately consumed. This behavior can be difficult or impossible to implement depending on the target domain. Variations in the token consumption can lead to drastic changes in the entire behavior of the activity. Therefore, it is important to include these variations in the semantics in order to accurately verify the system. This paper analyzes two variations that perform token transfer separately from token consumption (Figure \ref{fig:TokenC}). Variation 1 sequentially transfers all possible tokens from output pins to input pins before any node execution. Variation 2 performs both token transfer and node execution (if available) in a non-deterministic way.

In order to formally specify these variations, a \textit{micro-step} has to be added to the semantics, which is responsible for the token transfer, as is shown in the following rule. Furthermore, the token consumption of the \textit{macro-steps} of the reference semantics has to be limited to the input pins.

\begin{table}[!h] \begin{minipage}{0.98\linewidth}\centering \footnotesize
\begin{equation*}\frac{\parbox{4.7in} {
\vspace{-0.4cm}
$a \in A, e \in edge(a),S_a(a)=\langle executing, P_s, P_n\rangle, transfer(e) \neq \emptyset,$\\
$Vc=transfer(e),$\\
$V_s=S_{th}(source(e)),$\\
$V_t=S_{th}(target(e))$
}}{ \parbox{4.7in} {
$\langle  S_n, S_a,  S_{th}, S_{\Sigma} \rangle  \rightsquigarrow \langle S_n , S_a,  S_{th}[source(e) \mapsto V_s \rceil Vc][target(e) \mapsto V_t \cup Vc], S_{\Sigma} \rangle$
 }}
\end{equation*}
\end{minipage}
\begin{minipage}{0.01\linewidth}\scriptsize \begin{flushleft}
\vspace{-0.25cm} {\ttfamily \bfseries 1}\\
\vspace{0.11cm}	{\ttfamily \bfseries 2}\\
\vspace{0.11cm}	{\ttfamily \bfseries 3}\\
\vspace{0.11cm}	{\ttfamily \bfseries 4}\\
\vspace{0.3cm} {\ttfamily \bfseries 5}
\end{flushleft}\end{minipage}
\end{table}

Additionally for Variation 1, line 1 of the definition of a \textit{transition} has to be changed in order to ensure that all token transfers are performed before any execution takes place. The extended line is shown below.

\begin{equation*} 
\langle S_n, S_a,  S_{th}, S_{\Sigma} \rangle (\rightsquigarrow)^*  \langle S_n', S_a',  S_{th}', S_{\Sigma}' \rangle \not \rightsquigarrow, \langle S_n', S_a',  S_{th}', S_{\Sigma}' \rangle \twoheadrightarrow \langle S_n'', S_a'',  S_{th}'', S_{\Sigma}'' \rangle
\end{equation*}

A part of the state-space of the activity \textit{A} with different token consumption semantics is shown in figure \ref{fig:TokenC}. Variation 1 presents the most limited behavior, followed by the reference semantics. Note that the action \textit{C} is never executed in Variation 1 (denoted by (\textit{i(C)})). The reference semantics can finalize in two states, in which either \textit{C} or \textit{D} has been executed since both are competing for an outgoing token of \textit{A}. By contrast, Variation 2 has three final states. In two of them, \textit{D} is executed, one where the token between \textit{B} and \textit{C} has been transferred and one where not. Variation 2 enables more behaviors, since the position of the tokens is indirectly taken into account.

\section{Conclusion}

This paper has presented a reference operational semantics for UML 2.x activity diagrams.  The presented semantics aims to provide the same flexibility and extensibility as the standard, and is defined by inference rules that specify all allowed behaviors by the standard. These rules can be extended or additional rules can be added in order to customize the semantics to accurately specify characteristics of the system implementation and the domain requirements.  Since the semantics is operational, it also serves as a basis for translating activity diagrams into model-checking tools. The correctness of the semantics cannot be formally verified, since the standard does not provide a formal reference and there is not yet a consensus formal semantics for UML activity diagrams. Therefore, we also want with our extensible semantics to open a discussion in the community about a common formal framework for UML models.

For future work, we want to implement the generation of Kripke structures based on the presented inference rules, thereby facilitating model checking of activities by translating the resulting structures to model-checker languages. In addition, although the proposed semantics can specify parallelism between executions of multiple nodes, the start and termination of multiple executions are interleaved. In order to support full parallelism, we are studying methods to extend the step semantics using notions from partial-order reduction. These methods can be applied to the resulting Kripke structure, which has information about the dependency between elements caused by shared resources. 

\bibliographystyle{splncs03}
\bibliography{/Users/zdaw/Documents/postdoc/Literature/LiteratureBib.bib}

\begin{thebibliography}{10}
\providecommand{\url}[1]{\texttt{#1}}
\providecommand{\urlprefix}{URL }

\bibitem{Borger00b}
B\"{o}rger, E., Cavarra, A., Riccobene, E.: An ASM Semantics for UML Activity
  Diagrams, vol. 1816, pp. 293--308. Springer Berlin Heidelberg (2000)

\bibitem{Broy11}
Broy, M., Cengarle, M.: Uml formal semantics: lessons learned. Software \&
  Systems Modeling  10(4),  441--446 (2011)

\bibitem{Cao06}
Cao, H., Ying, S., Du, D.: Towards model-based verification of bpel with model
  checking. In: Computer and Information Technology, 2006. CIT '06. The Sixth
  IEEE International Conference on. pp. 190--190 (Sept 2006)

\bibitem{Daw2015b}
Daw, Z., Cleaveland, R.: An extensible operational semantics for uml activity
  diagrams. In: International Conference on Software Engineering and Formal
  Methods (2015)

\bibitem{Daw2016}
Daw, Z., Cleaveland, R.: A semantics-based framework for uml simulation and
  verification. In: Model Driven Engineering Languages and Systems (2016 -
  submitted)

\bibitem{Daw13}
Daw, Z., Cleaveland, R., Vetter, M.: Integrating of model checking and uml
  based model-driven development for embedded systems. In: Automated
  Verification of Critical Systems (AVOCS), Electronic Communications of the
  EASS (2013)

\bibitem{Daw2014}
Daw, Z., Cleaveland, R., Vetter, M.: Formal verification of software-based
  medical devices considering medical guidelines. International Journal of
  Computer Assisted Radiology and Surgery  9(1),  145--153 (2014)

\bibitem{Daw2009b}
Daw, Z., Vetter, M.: Deterministic UML Models for Interconnected Activities and
  State Machines, vol. 5795, pp. 556--570. Springer Berlin Heidelberg (2009)

\bibitem{Eshuis06}
Eshuis, R.: {Symbolic model checking of UML activity diagrams}. ACM Trans.
  Softw. Eng. Methodol.  15(1),  1--38 (2006)

\bibitem{Grobelna10}
Grobelna, I., Grobelny, M., Adamski, M.: Petri nets and activity diagrams in
  logic controller specification - transformation and verification. In: Mixed
  Design of Integrated Circuits and Systems (MIXDES), 2010 Proceedings of the
  17th International Conference. pp. 607--612 ({June} 2010)

\bibitem{Groenniger10}
Gr\"{o}nniger, H., Rei\ss, D., Rumpe, B.: Towards a semantics of activity
  diagrams with semantic variation points. In: MODELS. pp. 331--345.
  Springer-Verlag, Berlin, Heidelberg (2010)

\bibitem{Guelfi05}
Guelfi, N., Mammar, A.: A formal semantics of timed activity diagrams and its
  promela translation. In: Software Engineering Conference, 2005. APSEC '05.
  12th Asia-Pacific. pp. 8 pp.-- ({Dec} 2005)

\bibitem{Gulan13}
Gulan, S., Johr, S., Kretschmer, R., Rieger, S., Ditze, M.: Graphical modelling
  meets formal methods. In: Industrial Informatics (INDIN), 2013 11th IEEE
  International Conference on. pp. 716--721 (July 2013)

\bibitem{Knieke12}
Knieke, C., Schindler, B., Goltz, U., Rausch, A.: Defining domain specific
  operational semantics for activity diagrams. Tech. rep., TU Clausthal (2012)

\bibitem{Lano09}
Lano, K.: UML 2 semantics and applications. Wiley Online Library (2009)

\bibitem{UML241}
OMG: {Unified Modeling Language, Superstructure, Version 2.4.1}.
  http://www.omg.org/spec/UML/2.4.1/Superstructure/PDF (2011)

\bibitem{Staines08}
Staines, T.: Intuitive mapping of uml 2 activity diagrams into fundamental
  modeling concept petri net diagrams and colored petri nets. In: 15th Annual
  IEEE International Conference and Workshop on the Engineering of Computer
  Based Systems. pp. 191--200 (March 2008)

\bibitem{Storrle04}
St\"{o}rrle, H.: Structured nodes in uml 2.0 activities. Nordic Journal of
  Computing  11(3),  279--302 (2004)

\bibitem{Storrle05b}
St\"{o}rrle, H.: Semantics and verification of data flow in uml 2.0 activities.
  Electronic Notes in Theoretical Computer Science  127(4),  35--52 (2005)

\bibitem{Storrle05}
St\"{o}rrle, H., Hausmann, J.: Towards a formal semantics of uml 2.0
  activities. In: In Proceedings German Software Engineering Conference.
  vol.~64 (2005)

\bibitem{Lam07}
Vitus S.~W, L.: A formalism for reasoning about uml activity diagrams. In:
  Nordic Journal of Computing. vol.~14 (January 2007)

\bibitem{Xu08}
Xu, D., Liu, Z., Liu., W.: Towards formalizing uml activity diagrams in csp.
  In: International Symposium on Computer Science and Computational Technology.
  vol.~2 (2008)

\end{thebibliography}

\end{document}